\documentclass[10pt,lettersize,journal]{IEEEtran}
\usepackage{amsmath,amsfonts}
\usepackage{algorithmic}
\usepackage{algorithm}
\usepackage{array}
\usepackage[caption=false,font=normalsize,labelfont=sf,textfont=sf]{subfig}
\usepackage{textcomp}
\usepackage{stfloats}
\usepackage{url}
\usepackage{verbatim}
\usepackage{graphicx}
\usepackage{cite}
\usepackage{glossaries}
\usepackage[nolist]{acronym}
\usepackage[numbers, sort&compress]{natbib}

\usepackage{tikz}
\usetikzlibrary{shapes,arrows.meta,fit}
\usepackage[linesnumbered,ruled,vlined,algo2e]{algorithm2e}
\newcommand{\V}[1]{\textit{#1\/}}%new var

\usepackage{flexisym}
\usepackage{makecell}
\usepackage{pgfplots}
\pgfplotsset{compat=1.8}

\usepackage{multirow}

\usepackage[pagebackref=true,breaklinks=true,colorlinks=true,bookmarks=false]{hyperref}
\hypersetup{colorlinks, allcolors=., urlcolor=violet}

\usepackage{orcidlink}

\begin{document}
% \setcounter{page}{3224}
% \linenumbers
\title{Deep Active Audio Feature Learning in Resource-Constrained Environments}

\author{Md Mohaimenuzzaman\orcidlink{https://orcid.org/0000-0002-9798-8136}, Christoph Bergmeir\orcidlink{https://orcid.org/0000-0002-3665-9021} and Bernd Meyer\orcidlink{https://orcid.org/0000-0003-1080-0338}
        % <-this % stops a space
%\thanks{Manuscript received 30 August 2023; revised 13 April 2024 and 12 June 2024; accepted 15 June 2024. Date of publication 21 June 2024; date of current version 27 June 2024. 
\thanks{This work was supported by the Australian Research Council under Grant DE19010 0 045. The associate editor coordinating the review of this manuscript and approving it for publication was Prof. Isabel Barbancho. (Corresponding author: Md Mohaimenuzzaman.)}
\thanks{Md Mohaimenuzzaman and Bernd Meyer are with the Department of Data Science and AI, Monash University, Clayton, VIC 3800, Australia (e-mail: md.mohaimen@monash.edu).}% <-this % stops a space
\thanks{Christoph Bergmeir is with the Data Science and Computational Intelligence Andalusian Institute, DaSCI, Andalucía, Spain, also with the Department of Computer Science and AI, University of Granada, 18071 Granada, Spain, and also with the Department of Data Science and AI, Monash University, Clayton, VIC 3800, Australia.}
%\thanks{Digital Object Identiﬁer 10.1109/TASLP.2024.3416697}
}

\markboth{Published at: IEEE/ACM TRANSACTIONS ON AUDIO, SPEECH, AND LANGUAGE PROCESSING, VOL. 32, 2024}{}
%{MOHAIMENUZZAMAN \MakeLowercase{\textit{et al.}}: DEEP ACTIVE AUDIO FEATURE LEARNING IN RESOURCE-CONSTRAINED ENVIRONMENTS}

% Remember, if you use this, you must call \IEEEpubidadjcol in the second

\maketitle
\begin{acronym}[Abbreviations]
    \acro{ae}[AE]{Audio Event}
    \acro{afl}[AFL]{Active Feature Learning}
    \acro{aed}[AED]{Audio Event Detection}
    \acro{al}[AL]{Active Learning}
    \acro{alknc}[AL-KNC]{Active Learning with K-Neighbors Classifier}
    \acro{allgr}[AL-LReg]{Active Learning with Logistic Regression}
    \acro{alridgec}[AL-RidgeC]{Active Learning with Ridge Classifier}
    \acro{bac}[BaC]{Bioacoustic Classification}
	\acro{bar}[BaR]{Bioacoustic Recognition}	
	\acro{btf}[BTF]{Black-throated Finch}
	\acro{cnn}[CNN]{Convolutional Neural Network}
	\acro{crnn}[CRNN]{Convolutional Recurrent Neural Network}
	\acro{cv}[CV]{cross validation}
	\acro{dafl}[DAFL]{Deep Active Feature Learning}
	\acro{dcnn}[Deep-CNN]{Deep Convolutional Neural Network}
	\acro{dan}[DAN]{Deep Acoustic Network}
	\acro{dal}[DAL]{Deep Active Learning}
	\acro{dicl}[DIcL]{Deep Incremental Learning}
	\acro{dl}[DL]{Deep Learning}
	\acro{dnn}[DNN]{Deep Neural Network}
	\acro{esc}[ESC]{Environmental Sound Classification}
	\acro{esc10}[ESC-10]{ESC with 10 classes}
	\acro{esc50}[ESC-50]{ESC with 50 classes}
	\acro{esn}[ESN]{Echo State Network}
	\acro{fcnn}[FCNN]{Fully Connected Neural Network}
	\acro{flops}[FLOPs]{floating point operations}
	\acro{icl}[IcL]{Incremental Learning}
	\acro{iot}[IoT]{Internet of Things}
	\acro{iwbeat}[iWingBeat]{InsectWingBeat}
	\acro{kd}[KD]{Knowledge Distillation}
	\acro{knn}[K-NN]{K-Nearest Neighbors}
	\acro{lr}[LR]{Logistic Regression}
	\acro{mcu}[MCU]{Microcontroller Unit}
	\acro{mcus}[MCUs]{Microcontroller Units}
	\acro{ml}[ML]{Machine Learning}
	\acro{ndl}[Non-DL]{Non-Deep Learning}	
	\acro{nn}[NN]{Neural Network}
	\acro{rac}[RAC]{Raw Audio Classification}
	\acro{rar}[RAR]{Raw Audio Recognition}
	\acro{rdg}[Ridge]{Ridge}
	\acro{rf}[RF]{Random Forest}
	\acro{rnn}[RNN]{Recurrent Neural Network}
	\acro{sota}[SOTA]{state-of-the-art}
	\acro{sed}[SED]{Sound Event Detection}
	\acro{svm}[SVM]{Support Vector Machine}
	\acro{tda}[TDA]{Target Device Architecture}
	\acro{tnn}[TNN]{Transformer Neural Networks}
	\acro{ts}[TS]{Time Series}
	\acro{tsc}[TSC]{Time Series Classification}	
	\acro{us8k}[US8K]{UrbanSound8k}
	\acro{95ci}[95\%CI]{95\% Confidence Interval}
	\acro{ce}[CE]{Cross Entropy}
	\acro{kld}[KLD]{KL Divergence}
	\acro{sgd}[SGD]{Stochastic Gradient Descent}
	\acro{adam}[ADAM]{Adaptive Momentum Estimation}
    % \acro{ftune}[fine-tune]{Retrain a \ac{dnn} model with a small set of new data mixed with/without a similar amount of old data}
    % \acro{ftuning}[fine-tuning]{Retrain a \ac{dnn} model with a small set of new data mixed with/without a similar amount of old data}
    % \acro{freeze}[freeze]{Fine-tune all the layers except some layers}
    % \acro{nfreeze}[no-freeze]{Fine-tune all the layers of the network}
    % \acro{ffreeze}[fixed-freeze]{Fine-tune all layers except some specific layers}
    % \acro{sfreeze}[scheduled-freeze]{Gradually unfreeze layers from the end and fine-tune them}
    % \acro{retrain}[retrain]{Retrain a \ac{dnn} model with the full training set}
\end{acronym}
\newglossaryentry{ftune}{
    name=fine-tune,
    description={Retrain a \ac{dnn} model with a small set of new data mixed with/without a similar amount of old data}
}
\newglossaryentry{ftuning}{
    name=fine-tuning,
    description={Retrain a \ac{dnn} model with a small set of new data mixed with/without a similar amount of old data}
}
\newglossaryentry{freeze}{
    name=freeze,
    description={Fine-tune all the layers except some layers}
}
\newglossaryentry{nfreeze}{
    name=no-freeze,
    description={Fine-tune all the layers of the network}
}
\newglossaryentry{ffreeze}{
    name=fixed-freeze,
    description={Fine-tune all layers except some specific layers}
}
\newglossaryentry{sfreeze}{
    name=scheduled-freeze,
    description={Gradually unfreeze layers from the end and fine-tune them}
}
\newglossaryentry{retrain}{
    name=retrain,
    description={Retrain a \ac{dnn} model with the full training set}
}
% \printglossary

\begin{abstract}
The scarcity of labelled data makes training \ac{dnn} models in bioacoustic applications challenging. In typical bioacoustics applications, manually labelling the required amount of data can be prohibitively expensive. To effectively identify both new and current classes, DNN models must continue to learn new features from a modest amount of fresh data. \ac{al} is an approach that can help with this learning while requiring little labelling effort. Nevertheless, the use of fixed feature extraction approaches limits feature quality, resulting in underutilization of the benefits of \ac{al}. We describe an \ac{al} framework that addresses this issue by incorporating feature extraction into the \ac{al} loop and refining the feature extractor after each round of manual annotation. In addition, we use raw audio processing rather than spectrograms, which is a novel approach. Experiments reveal that the proposed \ac{al} framework requires 14.3\%, 66.7\%, and 47.4\% less labelling effort on benchmark audio datasets ESC-50, UrbanSound8k, and InsectWingBeat, respectively, for a large \ac{dnn} model and similar savings on a microcontroller-based counterpart. Furthermore, we showcase the practical relevance of our study by incorporating data from conservation biology projects. All codes are publicly available on GitHub.
\end{abstract}


\section{Introduction}
\acresetall
\glsresetall
\IEEEPARstart{D}{eep} Neural Network (DNN) 
models require a large amount of labelled data and long training times to extract high-quality hierarchical features~\cite{ren2021DALSurvey}, limiting their success to domains with abundant labelled data~\cite{ash2019BatchAL}. Our research focuses on bioacoustic applications in conservation biology.
In this domain, generating labelled training samples  usually requires the manual extraction of short segments (e.g.\ individual bird calls) from continuous recordings that cover days, weeks, or even longer periods. The relevant segments often  occur only relatively infrequently in these recordings. Acquiring sufficient amounts of high-quality labelled data at the start of a project is thus often impractical. Ideally, we would like to bootstrap the process of sample collection with a small number of labelled samples. A first (tentative) classifier trained on these can then be used to suggest further relevant samples which are screened by an expert and subsequently used to train an improved classifier. The process is then repeated until the required performance is reached. Indeed, this is often done in real-world applications in an ad-hoc fashion~\cite{teixeira2022Adhoc}. A more principled approach to keep the labelling effort acceptable using an incremental strategy is to employ \acf{al}, a semi-supervised machine learning technique \cite{han2016SemiAlSEC}. 

As an iterative learning method, \ac{al} is intended to speed up learning, especially when a large labelled dataset is unavailable for traditional supervised learning \cite{settles2012AL, ren2021DALSurvey, ash2019BatchAL}. \ac{al} algorithms combine intelligent acquisition functions to select samples that promise the best differential learning \cite{ash2019BatchAL} with specialized incremental training techniques~\cite{han2016SemiAlSEC, ash2019BatchAL}.

Here, we investigate the suitability of \ac{al} for bioacoustic classification. From a high-level perspective, bioacoustic classification usually proceeds in two steps: the first step extracts a vector of characteristic features from the audio, while the second step performs a classification based on this feature vector. While Active Learning has been used in (bio)acoustic applications before (see Section~\ref{sec:al_related_works}), previous works have only attempted to improve the classification phase based on fixed feature extraction methods~\cite{ash2019BatchAL,coleman2020ALSEC, hilasaca2021VisualAL,roy2001ExpectedErrorReduct, han2016SemiAlSEC, shuyang2017ALSECMedoidAL, qian2017ALSEC, kholghi2018ALSEC, shuyang2018ALSEC, qin2019LearntDictSET, wang2019ALSEC, ji2019DictAlSEC, wang2019ALSEC, shuyang2020ALSED, shi2020TLIncAL, ren2021DALSurvey}. When using \ac{dal}, a pre-trained model extracts features from samples during preprocessing as input to classification models. These data are then utilised to train and \gls{ftune} the classifier. Using humans in the loop for additional labelling and correcting misclassified samples, it \gls{ftune}s the classifier. However, to the best of our knowledge, the feature extractor employed in the preprocessing stage has never been improved in previous works on environmental sound classification.  As a result, feature quality remains constant. Poor feature quality can result in low classification accuracy for even the most powerful classifier(s). Hence, the primary advantage of the \ac{dal} approach is never fully realized.

We hypothesize that retraining the feature extraction approach during the iterative training loop will improve feature quality, resulting in increased classifier accuracy. Using active learning, the aim is to optimize the feature extractor on the misclassified set as well as the newly labeled set by the human expert. After that, the classifier is fine-tuned with the features extracted by the optimized feature extractor. This improves the performance of active learning and allows us to reduce the amount of training samples required. Feature extraction continues to improve during this process so that the potential of \ac{dal} can be fully realized.

Our experiments demonstrate the validity of this hypothesis using three standard benchmarks: \ac{esc50}~\cite{piczakESC50DatasetArchive}, \ac{us8k}~\cite{salamon2014Urbansound8k}, and \ac{iwbeat}~\cite{chen2014FlyingInsect}. Including the feature extraction in the active learning loop results in a  model that requires significantly less labelling effort than existing DAL methods. For the three datasets, our approach reduces the labelling effort required by 14.28\%, 66.67\%, and 50\%, respectively. 

We demonstrate the method's practical relevance beyond standard benchmarks by applying it to data from a real-world conservation project.

In previous work~\cite{mohaimen2021ACDNet}, we have shown that bioacoustic classification can be achieved by surprisingly small models that allow us to perform the recognition \emph{in-situ} in field recording units. We show that our proposed active learning approach functions independently of the network size and that its advantages carry over to such tiny networks.

\section{Current Literature}
\label{sec:al_related_works}
The fundamental issue our work is addressing is that, traditionally, in the \ac{al} process, the feature extraction process is static instead of being included in the incremental active learning loop. 

\subsection{Audio Feature Extraction Approaches}
A significant variety of different audio feature extraction strategies are in use. Arguably, the most widely used methods are based on spectrograms (e.g., Mel Frequency Cepstral Coefficients (MFCC)~\cite{shuyang2017ALSECMedoidAL} including its first and second-order derivatives~\cite{shuyang2018ALSEC}, Log Power Mel Spectrogram (LPMS)~\cite{coleman2020ALSEC}, Chromagram~\cite{coleman2020ALSEC}, statistics of MFCCs in each audio segment~\cite{shuyang2018ALSEC}, etc.).
% \marginpar{\textcolor{red}{\bf References are now added}.}

An alternative is the use of raw audio \ac{ts} as input to a CNN and to let the  feature extractor  be learned as part of the model  \cite{mohaimen2021ACDNet, mohaimen2021pruningvsxnor, tokozume2017Envnet, tokozume2017Envnet2, huang2018Aclnet}.

Pre-trained \ac{dnn} models such as VGGish \cite{hershey2017VGGish} for audio classification are also in use, as are mixtures of hand-crafted features along with automatic features from the time-frequency domain of the audio signal used in \cite{shi2020TLIncAL} and learned dictionary-based techniques such as Gabor Dictionary \cite{schroder2016GaborDict}. Speech recognition adds further feature extraction models, including Wav2Vec~\cite{schneider2019wav2vec}, as well as feature models utilised for knowledge distillation~\cite{you2021knowledge,you2021mrd}. Nevertheless, it is well understood that general acoustic event recognition has very different requirements from speech recognition so that models designed for speech recognition are not directly transferable~\cite{CNK09, Zh08}. Speech models are thus not widely used in general acoustic event classification. We have also found in previous studies focused on bioacoustic recognition that knowledge distillation appears to be less effective than presumably in speech recognition and that straightforward structured model compression appears to be preferable~\cite{mohaimen2021ACDNet, mohaimen2021pruningvsxnor, mohaimen2022thesis}. 

\subsection{Overview of Deep Active Learning}
The basic idea behind active learning is that if a \ac{ml} algorithm is allowed to choose which data to learn from, it may be able to improve its accuracy while using fewer training labels \cite{settles2012AL, ren2021DALSurvey}. Typically,  \ac{al} algorithms request human annotators to label the data instances for learning. 

The essential components of an \ac{al} technique are the selection of samples for human annotation, incremental training of a classifier with the annotated data, and labelling the remaining samples with the trained classifier \cite{settles2012AL, ren2021DALSurvey, ash2019BatchAL, han2016SemiAlSEC}. Alternatively, co-training can be employed to label the remaining samples, where humans label instances that are predicted with lower confidence by the trained classifier \cite{zhang2014cooperative, jones2003active}. This is an iterative process. The quantity of samples that can be manually annotated is usually assumed as a fixed labelling budget. 

As it is difficult to train a \ac{dl} model to achieve acceptable performance with small amounts of data, the majority of works from the literature (e.g., \cite{eyben2010opensmile, shuyang2017ALSECMedoidAL, coleman2020ALSEC, hilasaca2021VisualAL, kholghi2018ALSEC, wang2019ALSEC}) employ ML algorithms such as \ac{rf}~\cite{breiman2001RandomForest}, \ac{svm}~\cite{cortes1995SVM}, \ac{knn} Classifier~\cite{fix1989KNN}, and \ac{lr}~\cite{cox1958lgr}. 

% \cite{jones2003al}Paper 1: All it does is try to show if feature extraction is not even high quality, their approach of cotraining based active learning to retrain/fine-tune the target fuction can compensate it. Furthermore, this work is based on speech and language modeling. It does not consider the initial feature extraction technique that is used to convert the samples to be machine readable and put in unlabeled sample pool.

% \cite{vendrig2002TrecAl}Paper 2: This paper improves the model that selects sample with user defined semantic concepts and domain experts. This improves the model that selects samples by retraining the model with samples labeled by domain expert. The video sample after being labeled, are transformed using multimedia descriptors to use in the target function / classifier. This work is fundamentally different with what we are doing in our paper. Our argument is that the multimedia descriptors are never fine-tuned. We leverage this gap in our research.

There are a variety of sample selection techniques or acquisition functions available. Commonly used examples of such techniques are sample selection by domain experts~\cite{vendrig2002trec}, Random sample selection, uncertainty sampling \cite{lewis1994UncertaintySampling} (e.g., least confidence, margin of confidence, etc.), diversity sampling (e.g., medoid-based \ac{al} \cite{shuyang2017ALSECMedoidAL}, furthest traversal \cite{basu2004ALFarthestTraversal}, cluster-based outliers, model-based outliers, etc.), query-by-committee \cite{seung1992QueryCommittee}, explore and constrained clustering~\cite{yu2017ALConstrainedClustering}, sample selection by direct perceptual user feedback~\cite{nielsen2014perception} and expected error reduction \cite{roy2001ExpectedErrorReduct}. However, as our acquisition function, we employ the BADGE~\cite{ash2019BatchAL} sample selection technique, which takes predicted uncertainty and sample variety into account to pick the most problematic samples. This is the only approach that delivers satisfactory results on the datasets used in this study.  The availability of the implementation code in a public repository further ensures the reproducibility of this method.
% \marginpar{\textcolor{red}{Added as per Christoph's suggestion}}

\subsection{Related Works}
The majority of contemporary literature has been devoted to event classification. \citet{han2016SemiAlSEC} employ the open-source openSMILE \cite{eyben2010opensmile} toolkit to extract audio features from the FindSound \cite{findsounds} dataset, the least confidence sampling strategy for sample selection, and SVM with linear kernels to train and classify the audio events. The same feature extraction toolkit and classifier are utilized by \citet{qian2017ALSEC} in the \ac{al} system for the classification of bird sounds. They select samples using a random selection technique for human annotation and continue annotating until the budget runs out or the performance of the classifier is adequate.

\citet{shuyang2017ALSECMedoidAL} employ MFCC and its first and second-order derivatives to extract features from audio data. They further use the statistics of MFCCs in each segment: minimum, maximum, median, mean, variance, skewness, kurtosis, median, and variance of the first and second-order derivatives in \cite{shuyang2018ALSEC}. They cluster the data using k-medoid clustering and annotate the medoids of each cluster as local representatives. The label is then propagated throughout the cluster. Once the budget for annotation is exhausted, an SVM and an RF are trained to predict the labels in the former and the latter work, respectively. If there is a discrepancy between propagated and predicted labels, the labels are submitted to humans for correction.

\citet{coleman2020ALSEC} use MFCC, LPMS, and Chromagram for audio feature extraction from the \ac{esc50} \cite{piczak2015ESC} dataset, train SVM with a small amount of annotated data for prediction of the labels of other samples, and use the smallest confidence scores to select samples for further human annotation. \citet{hilasaca2021VisualAL} also extract features from spectrograms for soundscape ecology data, cluster the data using k-medoid clustering, and select samples from clusters using random, medoid (samples closest to the cluster centroid), contour (samples furthest from the centroid), and their combinations, for human annotation. They use the \ac{rf} classifier to predict the labels for the rest of the data.

\citet{kholghi2018ALSEC} use acoustic indices for feature extraction, k-means clustering, and hierarchical clustering algorithms to cluster the data, and randomly listen to sounds from each cluster for human annotation. They train RF using the annotated data and use it to predict the labels. 

\citet{ji2019DictAlSEC} and \citet{qin2019LearntDictSET} use a Gabor Dictionary \cite{schroder2016GaborDict} and a learned dictionary for their \ac{al} based audio classification task on the \ac{us8k} and \ac{esc50} datasets. Both use k-medoid clustering to cluster the data and manually label the medoid of each cluster and propagate the labels to the clusters.

\citet{ash2019BatchAL} and \citet{shi2020TLIncAL} retrain the model in every iteration of human annotation and use the final trained model for labelling at the end of the learning process. \citet{ash2019BatchAL} adopt diverse gradient embeddings and the k-means++ \cite{vassil2006kmeans++} seeding algorithm in their acquisition function (i.e., BADGE) for taking predictive uncertainty and sample diversity into account; however, the proposed method has only been evaluated on image data. \citet{shi2020TLIncAL}, on the other hand, used data with extremely low-frequency of 1kHz and manually engineered the features. They evaluated their approach on ECG recordings obtained from patients with atrial fibrillation. Unlike the previous studies, \citet{wang2019ALSEC} extracted features from the sonic sensor data using a pre-trained VGGish audio model \cite{hershey2017VGGish}. They selected samples using an uncertainty sampling technique (i.e., least confidence score) and trained the RF classifier to label the data after human annotation.

In contrast to audio event classification, \citet{kim2017ALSED, kim2018ALSED} utilize MFCC to extract acoustic features from the DCASE2015 \cite{dcase2015} dataset for audio event detection. They calculate the distance among the samples using the nearest neighbour algorithm, rank the samples nearest to the previously annotated sample as high, and choose them for further annotation by a human expert. \citet{shuyang2020ALSED} use the spectrogram of TUT Rare Sound Events 2017 \cite{dcase2017TUT} and TAU Spatial Sound Events 2019 \cite{dcase2019TAU} datasets \cite{shuyang2020ALSED} and change point detection to identify segments from the spectrograms that have events. The segments are  clustered using k-medoid clustering, and the medoid of each cluster is annotated. The samples for human annotation are picked using the mismatch-first-farthest traversal approach suggested in their earlier study \cite{shuyang2018ALSEC}. In order to detect and classify the rest of the samples, they train a \ac{dnn} model architecture presented in \cite{xu2018NetSED}.

According to the above discussion, k-medoid clustering~\cite{park2009kmedoid} and the Farthest Traversal~\cite{basu2004ALFarthestTraversal} appear to be the most popular sample selection techniques. However, due to their computational complexity, they do not scale to large amounts of data. 

Importantly, all the proposed \ac{al} techniques are based on fixed feature sets that do not change throughout the \ac{al} life cycle. When a pre-trained model is used to extract features, it is never refined. To the best of our knowledge, it has never been investigated whether the fixed feature set is flexible enough to accommodate the model's behaviour to time-varying data features. Our work aims to integrate the feature extractor into the active learning loop so that it can be refined with each iteration. To make learned feature extraction as flexible as possible, we start from a raw audio-based automatic feature extraction technique proposed by \citet{mohaimen2021ACDNet, mohaimen2021pruningvsxnor} for end-to-end audio event classification.

\section{Proposed Active Learning Framework}
\label{sec:al_proposed_dafl}
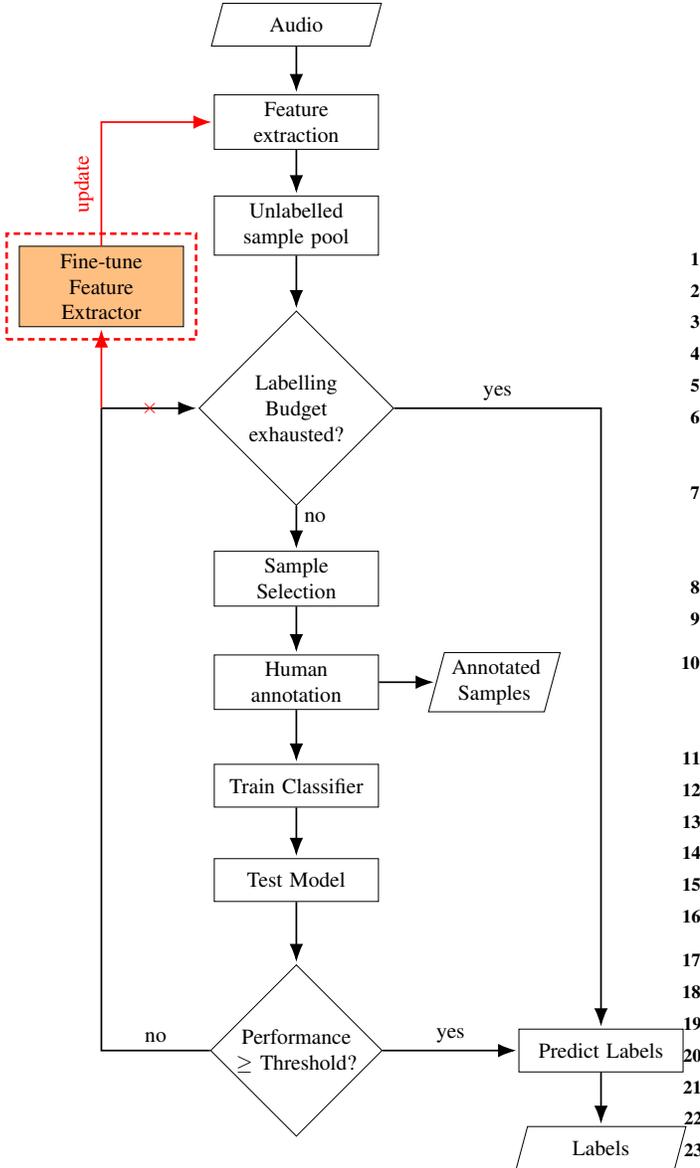
\begin{figure}[H]
    \hspace{-20px} 
	\scalebox{0.81}{%Project Flow Diagram
% Define block styles
\tikzstyle{decision} = [diamond, draw, fill=white, text width=6em, text badly centered, inner sep=0pt]
\tikzstyle{process} = [rectangle, draw, fill=white, text width=7em, text centered, minimum height=2em]
\tikzstyle{io} = [trapezium, draw, fill=white, text width=4em, text centered, minimum height=2em, trapezium left angle=75, trapezium right angle=105]
\tikzstyle{line} = [draw, thick, -{Latex[length=3mm]}, shorten >=1pt]
\tikzstyle{hbox} = [rectangle, draw, densely dashed, inner sep=2mm, very thick]

%\begin{center}
\begin{tikzpicture}
	%place nodes
	\node (input)[io]{Audio};
	\node (featureextraction)[process, below of=input, node distance=1.6cm]{Feature extraction};
	\node (sampelpool)[process, below of=featureextraction, node distance = 1.7cm] { Unlabelled sample pool};
	\node (checkbudget)[decision, below of =sampelpool, node distance=3cm] {Labelling Budget exhausted?};
	\node (sampleselection)[process, below of=checkbudget, node distance = 2.8cm] {Sample Selection};
	\node (annotate)[process, below of=sampleselection, node distance = 1.7cm] {Human annotation};
	\node (annotated)[io, right of=annotate, node distance = 3.25cm] {Annotated Samples};
	\node (trainclassifiers)[process, below of=annotate, node distance = 1.7cm] {Train Classifier};
	\node (testmodel)[process, below of=trainclassifiers, node distance=1.55cm] {Test Model};
	\node (performance)[decision, below of=testmodel, node distance=2.8cm] {Performance $\ge$ Threshold?};
	
	\node (labelling)[process,  right of=performance, node distance = 5cm] {Predict Labels};	
	\node (labels)[io, below of=labelling, node distance = 1.6cm] {Labels};
	
	\node (intersect) [rectangle, fill=white, left of = checkbudget, node distance=3.2cm]  {};
    \node (finetune)[process, fill=orange!50, above of=intersect, node distance=2cm] {Fine-tune Feature Extractor};
	\node (interest) [hbox, fit=(finetune), red]  {};

	% Draw lines
	\path [line] (input) -- (featureextraction);
	\path [line] (featureextraction) -- (sampelpool);
	\path [line] (sampelpool) -- (checkbudget);
	\path [line] (checkbudget)--node[near start, right] {no}(sampleselection);
	\path [line] (sampleselection) -- (annotate);
	\path [line] (annotate) -- (trainclassifiers);
	\path [line] (annotate) -- (annotated);
    \path [line] (trainclassifiers) -- (testmodel);
	\path [line] (testmodel) -- (performance);
	\path [line] (intersect.center) -- node[cross out, red]{$\large\times$}(checkbudget.west);
	\path [draw, thick, -,] (performance.west) -| node[near start, above]{no} (intersect.center);
	\path [line, red] (intersect.center) -- (finetune.south);
	
	\path [line, red] (finetune.north) |- node[near start, sloped, above] {update} (featureextraction.west);
	\path [line] (performance.east) -- node[midway, above]{yes} (labelling.west);
	\path [line] (labelling) -- (labels);
	\path [line] (checkbudget.east) -|node[near start, above]{yes}(labelling.north);
	
\end{tikzpicture}
%\end{center}
}
	\caption{The detailed architecture of the proposed \ac{dafl} where the red arrow indicates how the feature extractor is incorporated in the \ac{al} loop. Conventional systems use the black path instead of the red path for the active learning loop.}
	\label{fig:als_proposed_architecture}
    % \vspace{-30px}
\end{figure}

Figure~\ref{fig:als_proposed_architecture} depicts the detailed construction of the proposed \acf{dafl} approach.

\ac{dafl} allows us to \gls{ftune} the feature extractor. The output layer (i.e.\@ the final dense and softmax layers) is removed from the pre-trained model to accomplish this. 

The acquisition function is responsible for identifying the most informative, diverse, and challenging samples by integrating one or more sample selection techniques. We adopt the BADGE~\cite{ash2019BatchAL} sample selection technique in our acquisition function. It combines diverse gradient embeddings and the k-mean++~\cite{vassil2006kmeans++} seeding algorithm to take predictive uncertainty and sample diversity into consideration to identify the most difficult samples. The adoption of this sample selection technique in our acquisition function is motivated by the fact that it allows for the use of the k-means algorithm for medium to large datasets. K-means remains one of the most effective method used in sample selection techniques~\cite{park2009kmedoid,kim2017ALSED,kholghi2018ALSEC,kim2018ALSED,shuyang2020ALSED,hilasaca2021VisualAL}. Due to its computational complexity, namely with a time complexity of $\mathcal{O}(n^2)$, k-means does not scale well for large datasets~\cite{mohaimen2022thesis,fawaz2019DlForTsc}. BADGE incorporates k-means++ which uses a greedy approach for initial centroid selection to accelerate convergence. 

In our proposed method, the feature extractor is \gls{ftune}d in each iteration of human annotation (simulated) on the annotated examples to improve feature embeddings. The previously extracted features are then replaced with new features extracted from the \gls{ftune}d model. This iterative process is repeated until the labelling budget is exhausted or a desirable level of classification accuracy is attained. We then retrain the classifier(s) using the extracted features from the feature extractor. We present our proposed system in Algorithm~\ref{algo:deepfeatal}.

\begin{figure*}
    \centering
    \includegraphics[width=0.6\linewidth]{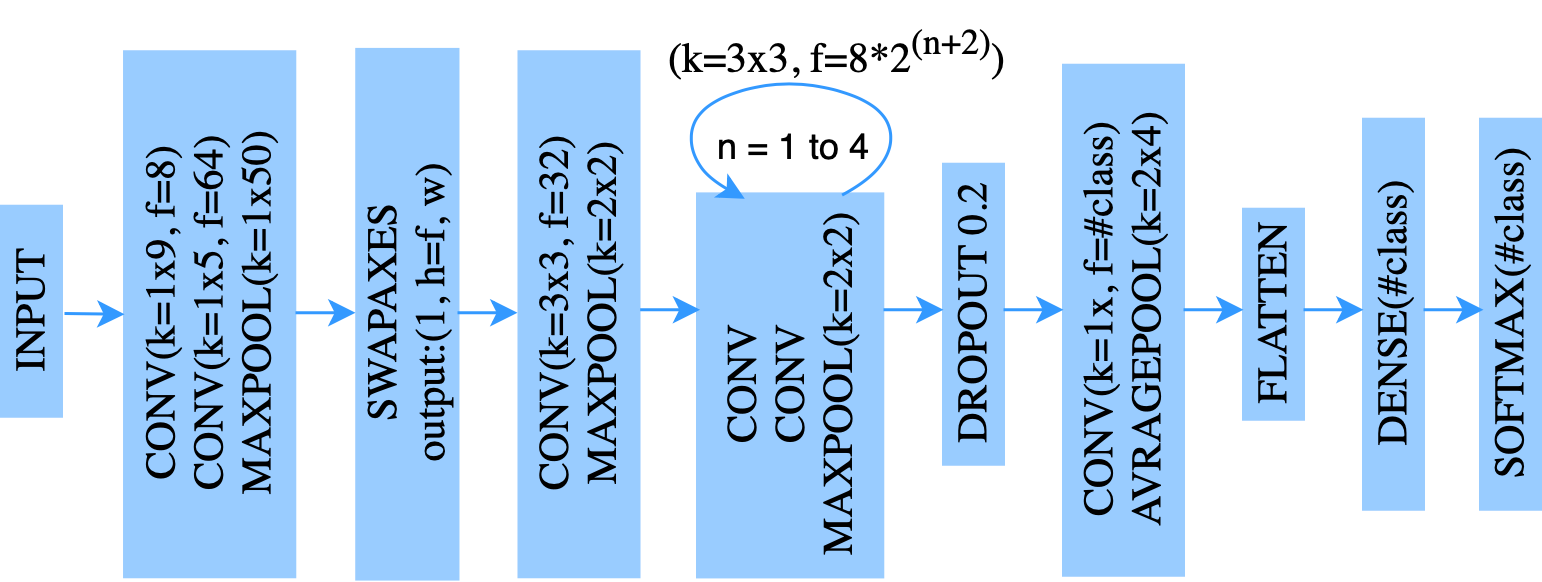}
    \vspace{-1pt}
    \caption{ACDNet architecture for an input length of 30,225, with $k$ and $f$ representing kernel size and number of filters, respectively, and $n \in \{1, \ldots, 4\}$. The height ($h$) and width ($w$) represent frequency and time resolution, respectively.}
    \label{fig:acdnet_full}
    \vspace{-10pt}
\end{figure*}

We employ ACDNet~\cite{mohaimen2021ACDNet} as the feature extractor. Figure~\ref{fig:acdnet_full} depicts the ACDNet architecture with an input length of 30,225.

\sloppy

\begin{algorithm}
    \DontPrintSemicolon
    \SetAlgorithmName{Algorithm}{algorithm}{DAFL}
    \SetNoFillComment
    \SetKwInput{KwInput}{Input}                % Set the Input
    \SetKwInput{KwOutput}{Output}    
    \caption{DAFL}
    \label{algo:deepfeatal}
    % \tcc{\textcolor{red}{Notations used in this procedure are self explanatory}}
    \KwInput{\ac{dnn} Model ($M$), Labelling Budget ($B$), \Gls{ftune} Epochs ($E$), Performance Threshold ($P_{thres}$), No. of Classes ($C$)}
    \KwOutput{Labeled Data ($L$)}
    \KwData{Labeled Set ($S_l$), Validation set ($S_v$), Test Set ($S_t$), Unlabeled Pool ($S_u$)}
    
    \BlankLine
    $isModelFound \longleftarrow \V{FALSE}$ \par
    $N \longleftarrow$ no. of samples to be selected \par
    $AF \longleftarrow$ Acquisition Function \par
    $S \longleftarrow$ Samples selected by $AF$ \par
    $HA \longleftarrow$ Human Annotation \par
    $C_l \longleftarrow$ $GetLayerCount(M)$ \par
    
    \BlankLine
    \tcc{last convolution index}
    $I_{lconv} \longleftarrow C_l-2$ \par
    \tcc{Replace the filters of the last convolution and train}
    $M \longleftarrow M.layers[I_{lconv}].filters \longleftarrow 5 * C )$ \par
    $M \longleftarrow Train(M)$ \par
    
    \BlankLine
    \While{$B > len(S_l)$ {\rm\bf{and}} $\V{NOT} \; isModelFound $}{
        \tcc{Keep layers up to the last convolution}
        $M^\prime \longleftarrow M - M[0 : I_{lconv}]$ \par
        $S \longleftarrow AF(M^\prime(S_u(N)))$ \par
        $S \longleftarrow HA(S)$ \par
        $S_u \longleftarrow S_u - S$ \par
        $S_l \longleftarrow S_l \cup S$ \par
        $S_{ft} \longleftarrow S_a \cup S_{ft}[$Random $len(S)$ samples$]$ \par
        
        \BlankLine
        $val \longleftarrow 0.0$ \par
        \ForEach{$e$ {\rm\bf{in}} $E$}{
            $M_f \longleftarrow$ \Gls{ftune} $M$ on $S_{ft}$ by minimizing $\V{KLD}$ loss \par
            $val^\prime \longleftarrow M_f(S_v)$ \par
            \If{$val^\prime > val$}{
                $val \longleftarrow val^\prime$ \par
                $M \longleftarrow M_f$
            }
        }
        
        \BlankLine
        $R_{es} \longleftarrow M(S_t)$ \par
        \If{$R_{es} \ge P_{thres}$}{
            $isModelFound \longleftarrow \V{TRUE}$ \par
        }
    }

    \BlankLine
    $L \longleftarrow M(S_u) \cup S_l$ \par
    
    \BlankLine
    {\rm\bf{return}} $L$ \par
\end{algorithm}
% \DecMargin{1em}
\normalsize

We have provided a comparison of different \ac{al} methods proposed for the \ac{us8k} dataset and our method in Table~\ref{tab:existing_al_vs_dafl_us8k}. However, it stands to reason that the very substantial performance differences to our method are not just due to the \ac{al} components but also due to the (much older and weaker performing) backbone networks used in these other approaches.

% \cite{shuyang2020ALSED}
\begin{table}[htb!]
    % \vspace{-5mm}
    \caption{Comparison between existing active learning methods and \ac{dafl} (proposed) on \ac{us8k} dataset.}
    \begin{center}
    \resizebox{0.9\linewidth}{!}{%
    \begin{tabular}{| c | c | c |} 
    	\hline
      	\thead{\acf{al} Method} & \thead{Labelling budget} & \thead{Accuracy}\\
      	\hline
      	\citet{shuyang2017ALSECMedoidAL} & 2,000 & 65.00\% \\
      	\hline
      	\citet{shuyang2018ALSEC} & 8,000 & 64.70\% \\
        \hline
      	\citet{qin2019LearntDictSET} & 8,000 & 67.50\% \\
      	\hline
      	\textbf{\ac{dafl} (proposed)} & 1,500 & 89.45\% \\
      	\hline
    \end{tabular}}
    % \vspace{-1mm}
    \label{tab:existing_al_vs_dafl_us8k}
  \end{center}
  % \vspace{-\baselineskip}
\end{table}

\section{Comparative Analysis}
\label{sec:al_comp_analysis}
In this section, we compare the performance of \ac{dafl} with the existing DAL techniques and a baseline technique that we refer to as \ac{dicl}. In \ac{dicl}, the samples are chosen at random for annotation. In contrast, the \ac{dal} and \ac{dafl} experiments use  the BADGE~\cite{ash2019BatchAL} sample selection technique,
and the model is \gls{ftune}d on the newly annotated samples. The implementation of BADGE is retrieved from \citet{ash2019BatchALCode} and used in our \ac{al} implementation. 
\ac{dicl} is thus the same as \ac{dafl} but without a targeted sample acquisition method. 

This study employs three standard benchmark datasets: \ac{esc50}, \ac{us8k} and \ac{iwbeat}. To determine the \gls{ftuning} strategy, a preliminary study was conducted. Experiments show that \gls{ftuning} the entire model produces the best results (see Section~\ref{sec:al_nfreeze_sfreeze_ffreeze}). We begin by contrasting the performance of \ac{dicl} and \ac{dal}. For the \ac{dal} experiments, we used three different classifiers available in scikit-learn~\cite{scikitLearn2011}: K-Neighbors Classifier, Logistic Regression, and Ridge Classifier. These methods are referred to as \ac{allgr}, \ac{alknc} and \ac{alridgec}, respectively. After analyzing the data, we choose the best method for each dataset and compare its performance to that of \ac{dafl} on each of the three datasets.

Our analysis shows that for \ac{esc50} and \ac{us8k}, \ac{alridgec} and \ac{allgr} outperform \ac{dicl}, whereas \ac{dicl} outperforms \ac{dal} for \ac{iwbeat}. We then compared the performance of these approaches to that of our proposed \ac{dafl} and discovered that \ac{dafl} performed significantly better. \ac{dafl} requires approximately 14.3\%, 66.67\%, and 47.4\% less labelling effort (according to Appendix~\ref{appendix_addtional_tables}~Table~\ref{tab:al_icl_vs_dal_dafl_esc50}, \ref{tab:al_icl_vs_dal_vs_dafl_us8k} and \ref{tab:al_icl_vs_dal_vs_dafl_iwingbeat}) than other techniques for \ac{esc50}, \ac{us8k} and \ac{iwbeat} datasets, respectively. The same set of experiments is performed with Micro-ACDNet, which is 97.22\% smaller than ACDNet and uses 97.28\% fewer \acf{flops}~\cite{mohaimen2021ACDNet}. \ac{dafl} outperforms all other approaches on all three datasets, requiring 42.85\%, 66.67\%, and 60\% less labelling effort, respectively. This supports the hypothesis we set out to investigate: including feature extraction in the \ac{al} loop improves performance and reduces labelling effort.

\subsection{NoFreeze vs. Freeze Layers for Deep Incremental Learning}
\label{sec:al_nfreeze_sfreeze_ffreeze}
% \marginpar{\textcolor{red}{Christoph suggested moving this under Section 3. Let's see what Bernd suggests.}}

\begin{figure}[H]
	\centering
    \begin{tikzpicture}
		\begin{axis}[legend pos=north east,
			width = 0.85\columnwidth,
			height = 7cm,
			xlabel=Learning Iterations,
			grid=major,
			ylabel=Accuracy (\%),
			xtick={0, 1, 2, 3, 4, 5, 6, 7},
            % symbolic x coords={Loop0,Loop1,Loop2,Loop3,Loop4,Loop5,Loop6,Loop7}, 
			ymin=55,ymax=66,
			ytick={56, 57, 58, 59, 60, 61, 62, 63, 64, 65},
			legend columns=3,
			ticklabel style = {font=\scriptsize},
  			label style = {font=\scriptsize},
			legend style={at={(0.5,-0.15)}, anchor=north,legend cell align=left,font=\scriptsize} %
			]
		\addplot[line width=1.25pt, color=red, mark=*] coordinates{
			(0, 55.75)(1, 59.50)(2, 61.25)(3, 61.75)(4, 64.00)(5, 63.50)
			(6, 64.75)(7, 65.50)
		};
		\addplot[line width=1.25pt, color=blue, mark=*] coordinates{
			(0, 55.75)(1, 58.00)(2, 59.25)(3, 59.75)(4, 62.00)(5, 62.50)
			(6, 62.75)(7, 63.50)
		};
		\addplot[line width=1.25pt, color=green, mark=*] coordinates{
			(0, 55.75)(1, 59.50)(2, 60.75)(3, 61.25)(4, 63.00)(5, 62.00)
			(6, 63.75)(7, 64.00)
		};
		\legend{\Gls{nfreeze}, \Gls{ffreeze}, \Gls{sfreeze}}
		\end{axis}
	\end{tikzpicture}
    \vspace{-5pt}
    \caption{Fine-tuning ACDNet in different settings (no-freeze, fixed-freeze and scheduled-freeze) for incremental learning. Please see Table~\ref{tab:al_acdnet_freeze_nofreeze_esc50} for the data used to generate this plot.}
	\label{plot:nofreeze_vs_freeze}
	% \vspace{-8pt}
\end{figure}
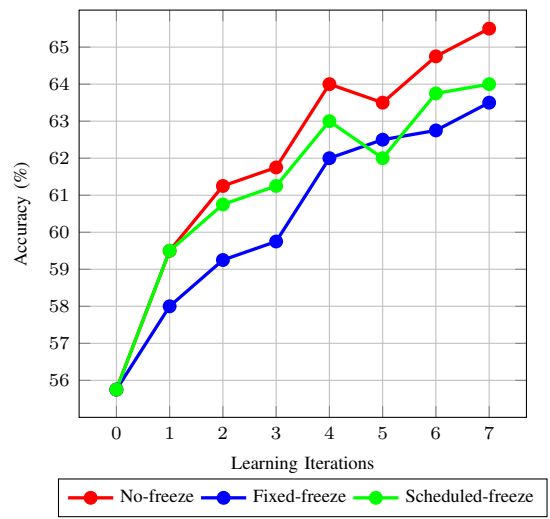

Our first experiment is designed to show whether the inclusion of feature extraction in the active learning loop is, in principle, useful and to select the best \gls{ftuning} method. To this end, we first train ACDNet on a part of the dataset and then \gls{ftune} it using three distinct strategies (see Appendix~\ref{sec:al_experimenal_details} for full training details).  
\begin{itemize}
\item \gls{nfreeze}: all ACDNet layers have the ability to \gls{ftune} their weight. 
\item \gls{ffreeze}: only the last three layers are allowed to \gls{ftune}. This is based on the notion that the initial layers perform feature extraction while the last three layers perform classification.
\item \gls{sfreeze}: the last three layers are unfrozen for a few epochs, then two more layers for a few more epochs, and finally two more layers for a few more epochs. 
\end{itemize}
This study was conducted using the \ac{esc50}~\cite{piczakESC50DatasetArchive} dataset. Experiment results show that \gls{ftuning} the entire network yields the best results (see Figure~\ref{plot:nofreeze_vs_freeze}). As a result, we will adopt the ``no-freeze'' strategy  as \gls{ftuning} for the remainder of this paper.

%Figure~\ref{plot:nofreeze_vs_freeze} (results also reported in Table~\ref{tab:al_acdnet_freeze_nofreeze_esc50}) shows that the model's performance is improved through the \gls{ftuning} process when the \gls{nfreeze} policy is followed. Although the improvement is modest, it is statistically significant. In the subsequent section of this article, we have shown that through statistical significance testing. This already gives clear evidence that it is useful to include fine-tuning the feature extraction layer in the active learning loop. As a result, we will adopt the ``no-freeze'' strategy  as \gls{ftuning} for the remainder of this paper.

%\subsection{Analysis with Full-sized ACDNet Model}
\subsection{Analysis with Full-sized Deep Neural Network Model}
\label{sec:al_comp_analysis_full_resource_computing}
For this analysis, the experiment is designed to select the best-performing conventional DAL strategy for our problem. The winning method will subsequently be compared with the proposed DAFL. 

As in \citet{mohaimen2021ACDNet}, we represent accuracy as a \ac{95ci}~\cite{carpenter2000bootstratCI4} calculated using bootstrap confidence intervals~\cite{diciccio1996bootstratCI1, cohen1995bootstratCI2}. We used bootstrap sampling with replacement to test the model 1,000 times on the test sets of all the datasets to calculate the \ac{95ci} (see Equation~\ref{eqn:estimated_95_ci} following \citet{mohaimen2021ACDNet}).
% \useshortskip
\begin{equation}
    95\%CI = \mu \pm Z \frac{\sigma}{\sqrt{N}}
    \label{eqn:estimated_95_ci}
\end{equation}
where $\mu$ is the average accuracy of all the tests, $Z=1.96$~\cite{carpenter2000bootstratCI4}, $\sigma$ is the standard deviation, and $N$ is the total number of tests.

%For \ac{dicl}, we select samples at random from the unlabeled pool and tune ACDNet. The method resembles Algorithm~\ref{algo:deepfeatal} excluding the acquisition function. 

Note that all samples used in this research have been labelled by human experts. We simulate the ``human annotation'' process by withholding the labels of data at the start of the experiment and placing the data in the unlabelled pool. We then deliver the withheld labels on demand as the acquisition picks these unlabelled samples, just as a human expert would. In the remainder of this paper, we refer to this as ``simulated human annotation''. While there is no human direct interaction with the process, this clearly delivers the exact same results. 

% \subsubsection{Deep Incremental Learning (DIcL) vs Existing Deep Active Learning (DAL) vs DAFL}

We first pitch \ac{dicl} against three conventional \ac{dal} strategies: AL-KNC, AL-LogisticReg and AL-RidgeC. We then compare the performance of the best method against our proposed \ac{dafl}.  In the case of \ac{dafl}, we use ACDNet as the learning and classification model rather than \ac{dal}. The primary distinction between the above methods and \ac{dafl} is that it is calibrated using annotated samples and the feature extractor is updated. We begin by comparing the performances of models derived from the above mentioned techniques, and to validate the findings, then we examine the statistical significance of the performance of the winning model. We present the analyses on the three datasets in the following order: \ac{esc50}, \ac{us8k}, and \ac{iwbeat}. 

To measure the statistical significance of each model's performance, we employ the statistical significance test described in~\cite{fawaz2019DlForTsc}. In particular, we use a Friedman test \cite{friedman1940FriedmanTest} to reject the null hypothesis of no difference within the whole group of methods. Then, in accordance with~\citet{benavoli2016ShouldWeUse}, we conduct a pairwise post-hoc analysis using the Wilcoxon signed-rank test~\cite{wilcoxon1992PostHoc} and Holm's correction~\cite{holm1979Holm, garcia2008Extension}, with an initial significance level of $\alpha=5\%$. As a graphical depiction, we use \citet{demvsar2006StatComp}'s Critical Difference (CD) diagram. In the CD diagram, a thick horizontal line shows that the differences in the accuracy of a set of classifiers are not statistically significant. The horizontal scale at the top indicates the ranks of the learning methods, and the numbers adjacent to them indicate the calculated score used to determine the rank of each method.

\subsubsection{Analysis on \ac{esc50} Dataset}
Figure~\ref{plot:fullsize_esc50_performance_analysis} which includes data from Appendix~\ref{appendix_addtional_tables}~Table~\ref{tab:al_icl_vs_dal_dafl_esc50}), shows that \ac{alridgec} achieves the highest prediction accuracy among \ac{dicl} (i.e., $65.43\%$) and \ac{dal} methods in only 3 iterations on the \ac{esc50} dataset, saving $\approx 57\%$ of the labelling budget. After seven simulated human labelling iterations, \ac{alridgec} obtains the highest accuracy of $67.94\%$, while our proposed method (\ac{dafl}) achieves the same accuracy in the sixth iteration ($68.74\%$), resulting in a 14.3\% reduction in the labelling budget. 

% \captionsetup[subfloat]{position=bottom, labelfont=scriptsize, textfont=scriptsize, justification=centering}
\begin{figure}[H]
	\centering
    % \subfloat[\ac{dicl} vs \ac{dal}]{
        \begin{tikzpicture}
    		\begin{axis}[legend pos=north east,
    			width = 0.85\columnwidth,
    			height = 7cm,
    			xlabel=Learning Iterations,
    			grid=major,
    			ylabel=\ac{95ci},
    			xtick={0,1,2,3,4,5,6,7},
                % symbolic x coords={0,Loop1,Loop2,Loop3,Loop4,Loop5,Loop6,Loop7}, 
    			ymin=53.5,ymax=70.5,
    			ytick={54, 56, 58, 60, 62, 64, 66, 68, 70},
    			legend columns=5,
    			ticklabel style = {font=\scriptsize},
      			label style = {font=\scriptsize},
    			legend style={at={(0.45,-0.15)}, anchor=north,legend cell align=left,font=\scriptsize} %
    			]
    		\addplot[line width=1.25pt, color=black, mark=*] coordinates{
    			(0, 55.81)(1, 59.56)(2, 61.23)(3, 61.71)(4, 63.94)(5, 63.38)
    			(6, 64.69)(7, 65.43)
    		};
    		\addplot[line width=1.25pt, color=blue, mark=*] coordinates{
                (0, 54.02)(1, 56.83)(2, 61.19)(3, 62.73)(4, 63.19)(5, 63.22)
                (6, 64.86)(7, 65.18)
    		};
    		\addplot[line width=1.25pt, color=brown, mark=*] coordinates{
    			(0, 58.78)(1, 60.76)(2, 63.19)(3, 63.81)(4, 64.04)(5, 65.46)
    			(6, 64.50)(7, 66.77)
    		};
    		\addplot[line width=1.25pt, color=green, mark=*] coordinates{
    			(0, 57.79)(1, 59.54)(2, 62.58)(3, 65.19)(4, 67.00)(5, 67.10)
    			(6, 67.73)(7, 67.94)
    		};
            \addplot[line width=1.25pt, color=red, mark=*] coordinates{
                (0, 55.81)(1, 63.76)(2, 64.15)(3, 65.00)(4, 65.94)(5, 67.28)
                (6, 68.74)(7, 70.02)
    		};
    		\legend{\ac{dicl}, \ac{alknc}, \ac{allgr}, \ac{alridgec}, \ac{dafl}}
    		\end{axis}
        \end{tikzpicture}
    \vspace{-5pt}
    \caption{\ac{dicl} vs \ac{dal} vs \ac{dafl} on \ac{esc50} where \ac{alknc}, \ac{allgr} and \ac{alridgec} are \ac{dal} methods.}
	\label{plot:fullsize_esc50_performance_analysis}
\end{figure}

\begin{figure}[H]
    \vspace{-15pt}
    \centering
    \includegraphics[width=1.0\linewidth]{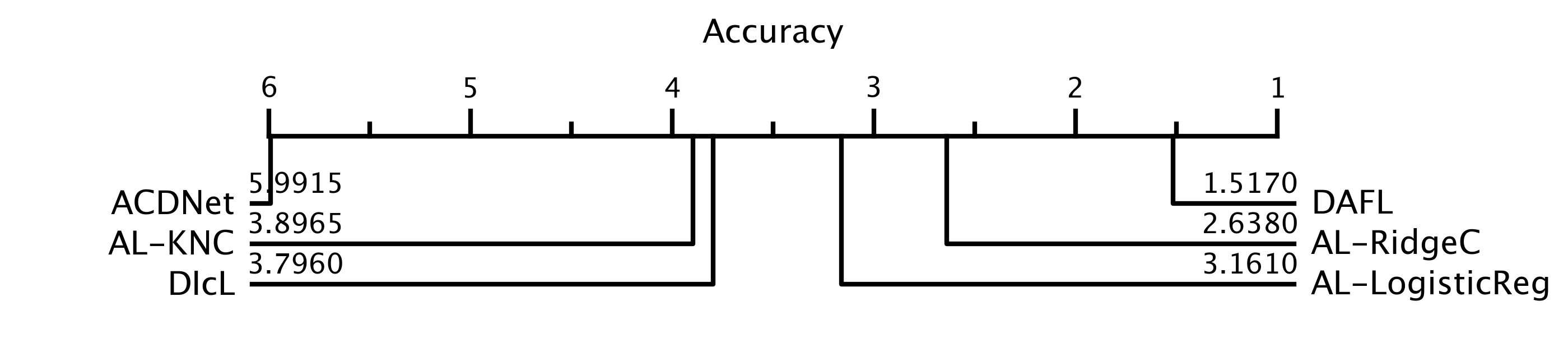}
    \vspace{-25pt}
    \caption{CD diagram showing statistical significance of the learning methods on the \ac{esc50} dataset.}
    \label{fig:esc50_sig_diag}
    \vspace{-8pt}
\end{figure}

Figure~\ref{fig:esc50_sig_diag} depicts the statistical significance of the final models produced after completing all human annotation iterations on the \ac{esc50} dataset for \ac{dicl}, \ac{dal}, and \ac{dafl}. According to the figure, all the models perform statistically significantly differently. It demonstrates that the model produced by \ac{dafl} outperforms other models, ranking top among them.

\subsubsection{Analysis on \ac{us8k} Dataset}
The same comparative analysis is performed on the \ac{us8k} dataset. Figure~\ref{plot:fullsize_us8k_performance_analysis} (refer to data in Appendix~\ref{appendix_addtional_tables}~Table~\ref{tab:al_icl_vs_dal_vs_dafl_us8k}) demonstrates that \ac{alknc} and \ac{allgr} exhibit significantly superior predictive capabilities compared to \ac{dicl}. Moreover, \ac{allgr} obtains the maximum prediction accuracy of \ac{dicl} (i.e., $88.19\%$), resulting in a savings of around $86.67\%$ of the labelling budget. As \ac{us8k} contains more data than \ac{esc50}, we perform fifteen iterations of human labelling for \ac{dicl} and the \ac{dal} process where \ac{allgr} achieves the highest prediction accuracy ($88.48\%$). Having said that, when we look at \ac{dafl}, it achieves that accuracy ($88.59\%$) in just five iterations, saving $66.67\%$ of the labelling budget.

\begin{figure}[H]
	\centering
    % \subfloat[\ac{dicl} vs \ac{dal}]{
        \begin{tikzpicture}
    		\begin{axis}[legend pos=north east,
    			width = 0.99\columnwidth,
    			height = 6cm,
    			xlabel=Learning Iterations,
    			grid=major,
    			ylabel=\ac{95ci},
    			xtick={0,1,2,3,4,5,6,7,8,9,10,11,12,13,14,15},
                % symbolic x coords={0,Loop1,Loop2,Loop3,Loop4,Loop5,Loop6,Loop7}, 
    			ymin=85.75,ymax=89.75,
    			ytick={86.0, 86.5, 87.0, 87.5, 88.0, 88.5, 89.0, 89.5},
                yticklabel style={/pgf/number format/.cd,fixed,fixed zerofill,precision=1},
    			legend columns=5,
    			ticklabel style = {font=\scriptsize},
      			label style = {font=\scriptsize},
    			legend style={at={(0.45,-0.19)}, anchor=north,legend cell align=left,font=\scriptsize} %
    			]
    		\addplot[line width=1.25pt, color=black, mark=*] coordinates{
        		(0, 86.97)(1, 87.12)(2, 86.94)(3, 87.47)(4, 87.46)(5, 87.98)
                (6, 87.80)(7, 87.78)(8, 87.94)(9, 86.91)(10, 87.67)
                (11, 88.19)(12, 87.67)(13, 87.00)(14, 86.71)(15, 87.38)
    		};
    		\addplot[line width=1.25pt, color=blue, mark=*] coordinates{
                (0, 87.24)(1, 87.60)(2, 87.51)(3, 87.49)(4, 87.66)(5, 87.65)
                (6, 88.05)(7, 88.14)(8, 88.12)(9, 88.04)(10, 87.93)
                (11, 88.05)(12, 87.91)(13, 87.80)(14, 88.18)(15, 88.48)
    		};
    		\addplot[line width=1.25pt, color=brown, mark=*] coordinates{
    			(0, 87.55)(1, 87.48)(2, 88.25)(3, 88.08)(4, 88.00)(5, 87.96)
                (6, 87.96)(7, 87.79)(8, 88.08)(9, 88.16)(10, 88.25)
                (11, 88.18)(12, 88.21)(13, 88.48)(14, 88.17)(15, 88.31)
    		};
    		\addplot[line width=1.25pt, color=green, mark=*] coordinates{
    			(0, 87.00)(1, 86.80)(2, 87.09)(3, 87.60)(4, 87.42)(5, 87.27)
                (6, 87.38)(7, 87.47)(8, 87.54)(9, 87.49)(10, 87.72)
                (11, 87.70)(12, 87.73)(13, 87.80)(14, 87.78)(15, 87.82)
    		};
            \addplot[line width=1.25pt, color=red, mark=*] coordinates{
                (0, 86.97)(1, 85.93)(2, 86.82)(3, 87.15)(4, 87.09)(5, 88.59)
                (6, 88.20)(7, 88.94)(8, 88.62)(9, 88.60)(10, 88.70)
                (11, 88.75)(12, 88.80)(13, 89.01)(14, 89.20)(15, 89.45)
            };
    		\legend{\ac{dicl}, \ac{alknc}, \ac{allgr}, \ac{alridgec}, \ac{dafl}}
            \end{axis}
        \end{tikzpicture}
        \vspace{-5pt}
    \caption{\ac{dicl} vs \ac{dal} vs \ac{dafl} on \ac{us8k} where \ac{alknc}, \ac{allgr} and \ac{alridgec} are \ac{dal} methods.}
    \label{plot:fullsize_us8k_performance_analysis}
\end{figure}
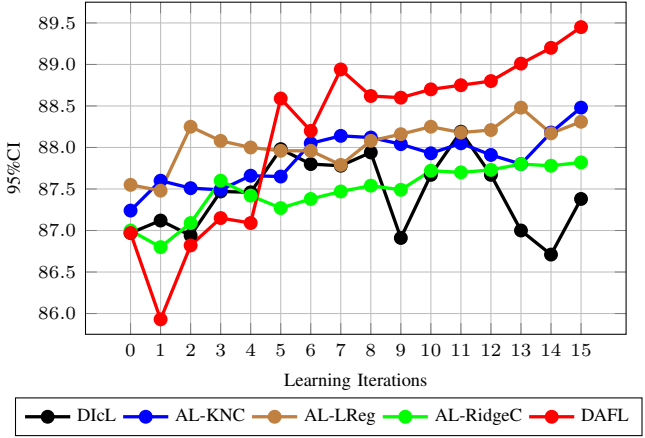

\begin{figure}[H]
    \vspace{-10pt}
    \centering
    \includegraphics[width=1.0\linewidth]{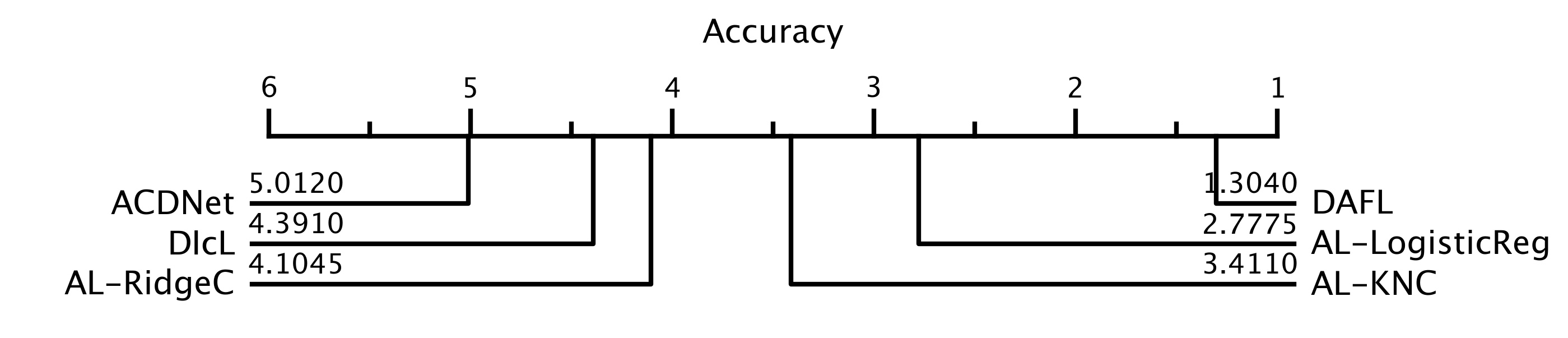}
    \vspace{-25pt}
    \caption{CD diagram showing statistical significance of the learning methods on the \ac{us8k} dataset.}
    \label{fig:us8k_sig_diag}
    \vspace{-8pt}
\end{figure}

An analogous statistical analysis is shown in Figure~\ref{fig:us8k_sig_diag} for the \ac{us8k} dataset. It illustrates that the model produced by \ac{dafl} outperforms other models and ranks highest among them.

\subsubsection{Analysis on \ac{iwbeat} Dataset}
Somewhat surprisingly, Figure~\ref{plot:iwingbeat_performance_analysis} reveals that \ac{dicl} has the highest prediction accuracy on the \ac{iwbeat} dataset. In contrast to the results on the other datasets, the performance of the initial model and the model obtained after human annotations (simulated) and \gls{ftuning} do not differ significantly for this dataset. Since this dataset contains more labelled samples than the previous two, we run twenty human annotation iterations. 

Figure~\ref{plot:iwingbeat_performance_analysis} illustrates a comparison of the performance of \ac{dicl}, \ac{dal} and our proposed \ac{dafl} methods on the \ac{iwbeat} dataset (data reported in Appendix~\ref{appendix_addtional_tables}~Table~\ref{tab:al_icl_vs_dal_vs_dafl_iwingbeat}). \ac{dicl} achieves the highest prediction accuracy ($68.86\%$) in the thirteenth iteration, while \ac{dafl} achieves that accuracy ($68.86\%$) in only five iterations, saving $47.4\%$ of the labelling budget.

\begin{figure}[ht]
	\centering
    % \subfloat[\ac{dicl} vs \ac{dal}]{
        \begin{tikzpicture}
    		\begin{axis}[legend pos=north east,
    			width = 0.99\columnwidth,
    			height = 6cm,
    			xlabel=Learning Iterations,
    			grid=major,
    			ylabel=\ac{95ci},
    			xtick={0,1,2,3,4,5,6,7,8,9,10,11,12,13,14,15,16,17,18,19,20},
                % symbolic x coords={0,Loop1,Loop2,Loop3,Loop4,Loop5,Loop6,Loop7}, 
    			ymin=65.5,ymax=70.5,
    			ytick={65.5, 66, 66.5, 67, 67.5, 68, 68.5, 69, 69.5, 70},
                yticklabel style={/pgf/number format/.cd,fixed,fixed zerofill,precision=1},
    			legend columns=5,
    			ticklabel style = {font=\scriptsize},
      			label style = {font=\scriptsize},
    			legend style={at={(0.45,-0.19)}, anchor=north,legend cell align=left,font=\scriptsize} %
    			]
    		\addplot[line width=1.25pt, color=black, mark=*] coordinates{
        		(0, 66.08)(1, 66.94)(2, 66.50)(3, 66.21)(4, 66.63)(5, 67.19)
                (6, 67.30)(7, 67.79)(8, 67.11)(9, 67.32)(10, 67.20)
                (11, 66.97)(12, 67.07)(13, 67.27)(14, 66.66)(15, 67.54)
                (16, 67.50)(17, 67.65)(18, 67.62)(19, 68.27)(20, 68.16)
    		};
    		\addplot[line width=1.25pt, color=blue, mark=*] coordinates{
                (0, 66.31)(1, 66.31)(2, 66.47)(3, 66.61)(4, 66.44)(5, 66.22)
                (6, 66.17)(7, 66.19)(8, 66.29)(9, 66.28)(10, 66.14)
                (11, 66.24)(12, 66.15)(13, 66.22)(14, 66.35)(15, 66.35)
                (16, 66.43)(17, 66.10)(18, 66.11)(19, 66.03)(20, 65.87)
    		};
    		\addplot[line width=1.25pt, color=brown, mark=*] coordinates{
    			(0, 66.11)(1, 66.53)(2, 66.57)(3, 66.53)(4, 66.64)(5, 66.61)
                (6, 66.68)(7, 66.79)(8, 66.91)(9, 66.95)(10, 67.02)
                (11, 67.15)(12, 67.05)(13, 67.10)(14, 67.10)(15, 67.03)
                (16, 67.13)(17, 67.17)(18, 67.17)(19, 67.11)(20, 67.17)
    		};
    		\addplot[line width=1.25pt, color=green, mark=*] coordinates{
    			(0, 65.82)(1, 65.79)(2, 65.95)(3, 66.02)(4, 65.89)(5, 65.87)
                (6, 66.03)(7, 66.01)(8, 66.14)(9, 66.04)(10, 66.16)
                (11, 66.12)(12, 66.36)(13, 66.32)(14, 66.20)(15, 66.14)
                (16, 66.25)(17, 66.31)(18, 66.37)(19, 66.30)(20, 66.28)
    		};
            \addplot[line width=1.25pt, color=red, mark=*] coordinates{
    			(0, 66.08)(1, 66.59)(2, 66.99)(3, 67.57)(4, 67.13)(5, 67.40)
                (6, 67.76)(7, 67.66)(8, 68.1)(9, 68.03)(10, 68.86)
                (11, 68.95)(12, 68.9)(13, 69.05)(14, 69.29)(15, 69.32)
                (16, 69.76)(17, 70.12)(18, 70.04)(19, 70.18)(20, 70.13)
    		};
    		\legend{\ac{dicl}, \ac{alknc}, \ac{allgr}, \ac{alridgec}, \ac{dafl}}
    		\end{axis}
	   \end{tikzpicture}
    \vspace{-5pt}
    \caption{\ac{dicl} vs \ac{dal} vs \ac{dafl} on \ac{iwbeat} where \ac{alknc}, \ac{allgr} and \ac{alridgec} are \ac{dal} methods.}
	\label{plot:iwingbeat_performance_analysis}
\end{figure}

\begin{figure}[H]
    \vspace{-10pt}
    \centering
    \includegraphics[width=1.0\linewidth]{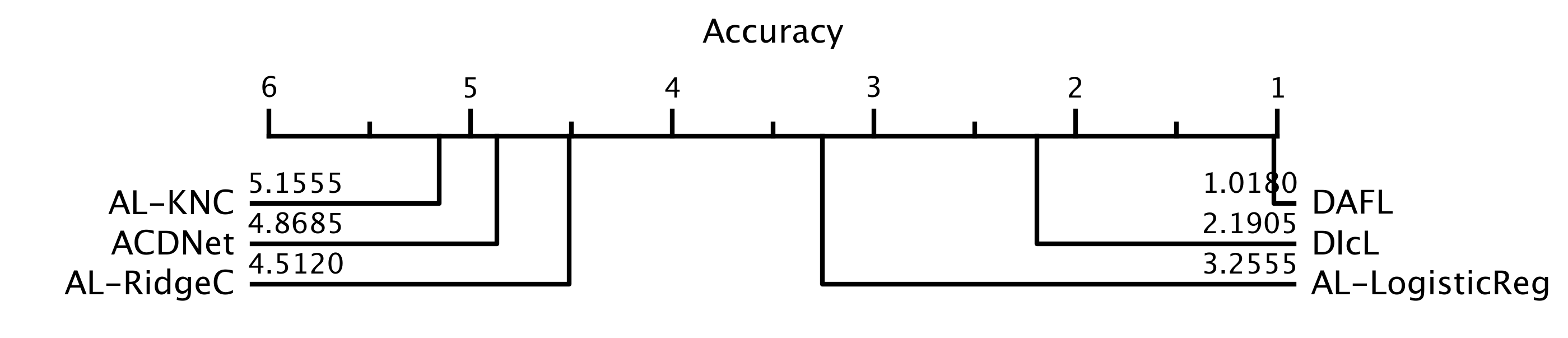}
    \vspace{-25pt}
    \caption{CD diagram showing statistical significance of the learning methods on \ac{iwbeat} dataset.}
    \label{fig:iwingbeat_sig_diag}
    \vspace{-8pt}
\end{figure}

When examining the statistical significance, the situation is the same regarding the \ac{iwbeat} dataset. Figure~\ref{fig:iwingbeat_sig_diag} shows that the \ac{dafl}-generated model outperforms other models and ranks first among them.

From the above comparative study, we observe that \ac{dafl} significantly outperforms the best-performing methods of \ac{dicl} and \ac{dal}, both in terms of final accuracy for a given labelling budget as well as in terms of labelling budget required for a given target performance. On the \ac{esc50}, \ac{us8k}, and \ac{iwbeat} datasets, \ac{dafl} requires approximately 14.3\%, 66.67\%, and 47.4\% less labelling effort than its counterparts, respectively. 

\subsection{Analysis with Edge-based Model}
\label{sec:al_acdnet_to_micro_acdnet}
In the preceding section, we illustrated the benefit of \ac{dafl} for ACDNet, a large architecture designed to run on performant hardware. Now, we repeat the study with Micro-ACDNet~\cite{mohaimen2021ACDNet}, an extremely small network operating on \ac{mcus}, to investigate if the benefits are retained when moving to very small models. Table~\ref{tab:al_micro_icl_vs_dal_vs_dafl_esc50}, \ref{tab:al_micro_icl_vs_dal_vs_dafl_us8k} and \ref{tab:al_micro_icl_vs_dal_vs_dafl_iwingbeat} summarizes the iteration-wise performances of \ac{dicl}, \ac{dal} and \ac{dafl} in terms of the \ac{95ci} for the \ac{esc50}, \ac{us8k}, and \ac{iwbeat} datasets, respectively.

Figures~\ref{plot:micro_esc50_performance_analysis}, \ref{plot:micro_us8k_performance_analysis} and \ref{plot:micro_iwingbeat_performance_analysis} use the data reported in Appendix~\ref{appendix_addtional_tables}~Table~\ref{tab:al_micro_icl_vs_dal_vs_dafl_esc50}, \ref{tab:al_micro_icl_vs_dal_vs_dafl_us8k} and \ref{tab:al_micro_icl_vs_dal_vs_dafl_iwingbeat}. Figure~\ref{plot:micro_esc50_performance_analysis} indicates that on the \ac{esc50} dataset, \ac{dafl} consistently outperforms \ac{dicl} and \ac{dal} (i.e., \ac{allgr}) by a significant margin. In fact, \ac{dafl} achieves the highest accuracy produced by \ac{allgr} (namely 63.42\% in seven iterations) in five iterations. As a result, 42.85\% of the labelling effort is saved. According to Figure~\ref{plot:micro_us8k_performance_analysis}, \ac{allgr} achieves 83.38\% accuracy in twelve iterations on the US8K dataset, whereas \ac{dafl} achieves 83.56\% in only five iterations. As a result, the amount of labelling effort required is reduced by 58.33\%. Figure~\ref{plot:micro_iwingbeat_performance_analysis} demonstrates that \ac{dafl} achieves the highest \ac{dicl} prediction accuracy (i.e. 63.88\% in eleven iterations) in only eight iterations on the \ac{iwbeat} dataset. This results in a 27.27\% decrease in labelling effort.
\begin{figure}[H]
	\centering
    \begin{tikzpicture}
    		\begin{axis}[legend pos=north east,
    			width = 0.85\columnwidth,
    			height = 6.5cm,
    			xlabel=Learning Iterations,
    			grid=major,
    			ylabel=\ac{95ci},
    			xtick={0,1,2,3,4,5,6,7},
                % symbolic x coords={0,Loop1,Loop2,Loop3,Loop4,Loop5,Loop6,Loop7}, 
    			ymin=47,ymax=66,
    			ytick={48, 50, 52, 54, 56, 58, 60, 62, 64, 66},
    			legend columns=5,
    			ticklabel style = {font=\scriptsize},
      			label style = {font=\scriptsize},
    			legend style={at={(0.45,-0.16)}, anchor=north,legend cell align=left,font=\scriptsize} %
    			]
            
    		\addplot[line width=1.25pt, color=black, mark=*] coordinates{
    			(0, 55.36)(1, 55.99)(2, 55.72)(3, 55.24)(4, 56.25)(5, 54.94)
    			(6, 57.37)(7, 57.70)
    		};
    		\addplot[line width=1.25pt, color=blue, mark=*] coordinates{
                (0, 47.89)(1, 48.89)(2, 53.80)(3, 55.35)(4, 58.83)(5, 57.37)
                (6, 57.92)(7, 56.67)
    		};
          	\addplot[line width=1.25pt, color=brown, mark=*] coordinates{
        		(0, 52.67)(1, 56.08)(2, 56.47)(3, 60.25)(4, 63.42)(5, 62.76)
                (6, 62.70)(7, 63.32)
    		};
    		\addplot[line width=1.25pt, color=green, mark=*] coordinates{
    			(0, 52.46)(1, 53.89)(2, 58.09)(3, 61.53)(4, 61.16)(5, 61.20)
    			(6, 60.70)(7, 62.38)
    		};
            \addplot[line width=1.25pt, color=red, mark=*] coordinates{
                (0, 55.36)(1, 58.31)(2, 60.05)(3, 60.76)(4, 63.61)(5, 63.25)
                (6, 64.01)(7, 65.20)
            };
    		\legend{\ac{dicl}, \ac{alknc}, \ac{allgr}, \ac{alridgec}, \ac{dafl}}
    		\end{axis}
        \end{tikzpicture}
    \vspace{-5pt}
	\caption{Edge-based \ac{dicl} vs \ac{al} vs \ac{dafl} on \ac{esc50} dataset}
	\label{plot:micro_esc50_performance_analysis}
	% \vspace{-8pt}
\end{figure}
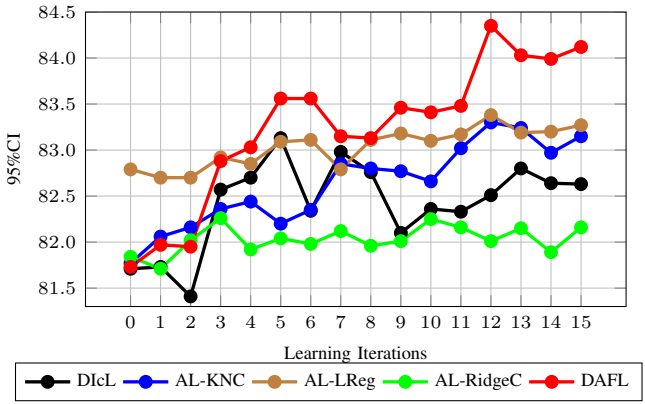
\begin{figure}[H]
	\centering
    \begin{tikzpicture}
        \begin{axis}[legend pos=north east,
            width = 0.99\columnwidth,
            height = 5.5cm,
            xlabel=Learning Iterations,
            grid=major,
            ylabel=\ac{95ci},
            xtick={0,1,2,3,4,5,6,7,8,9,10,11,12,13,14,15},
            % symbolic x coords={0,Loop1,Loop2,Loop3,Loop4,Loop5,Loop6,Loop7}, 
            ymin=81.3,ymax=84.5,
            ytick={81.5, 82.0, 82.5, 83.0, 83.5, 84.0, 84.5},
            yticklabel style={/pgf/number format/.cd,fixed,fixed zerofill,precision=1},
            legend columns=5,
            ticklabel style = {font=\scriptsize},
            label style = {font=\scriptsize},
            legend style={at={(0.45,-0.19)}, anchor=north,legend cell align=left,font=\scriptsize} %
            ]
    		\addplot[line width=1.25pt, color=black, mark=*] coordinates{
    			(0, 81.71)(1, 81.73)(2, 81.41)(3, 82.57)(4, 82.70)(5, 83.13)
                (6, 82.34)(7, 82.98)(8, 82.76)(9, 82.10)(10, 82.36)
                (11, 82.33)(12, 82.51)(13, 82.80)(14, 82.64)(15, 82.63)
    		};
    		\addplot[line width=1.25pt, color=blue, mark=*] coordinates{
                (0, 81.77)(1, 82.06)(2, 82.16)(3, 82.36)(4, 82.44)(5, 82.20)
                (6, 82.35)(7, 82.85)(8, 82.80)(9, 82.77)(10, 82.66)
                (11, 83.02)(12, 83.30)(13, 83.24)(14, 82.97)(15, 83.15)
    		};
          	\addplot[line width=1.25pt, color=brown, mark=*] coordinates{
    			(0, 82.79)(1, 82.70)(2, 82.70)(3, 82.92)(4, 82.85)(5, 83.09)
                (6, 83.11)(7, 82.79)(8, 83.11)(9, 83.18)(10, 83.10)
                (11, 83.17)(12, 83.38)(13, 83.19)(14, 83.20)(15, 83.27)
    		};
    		\addplot[line width=1.25pt, color=green, mark=*] coordinates{
    			(0, 81.84)(1, 81.71)(2, 82.02)(3, 82.26)(4, 81.92)(5, 82.04)
                (6, 81.98)(7, 82.12)(8, 81.96)(9, 82.01)(10, 82.25)
                (11, 82.16)(12, 82.01)(13, 82.15)(14, 81.89)(15, 82.16)
    		};
            \addplot[line width=1.25pt, color=red, mark=*] coordinates{
                (0, 81.73)(1, 81.97)(2, 81.95)(3, 82.88)(4, 83.03)(5, 83.56)
                (6, 83.56)(7, 83.15)(8, 83.13)(9, 83.46)(10, 83.41)
                (11, 83.48)(12, 84.35)(13, 84.03)(14, 83.99)(15, 84.12)
    		};
    		\legend{\ac{dicl}, \ac{alknc}, \ac{allgr}, \ac{alridgec}, \ac{dafl}}
    		\end{axis}
        \end{tikzpicture}
    \vspace{-5pt}
	\caption{Edge-based \ac{dicl} vs \ac{dal} vs \ac{dafl} on \ac{us8k} dataset}
	\label{plot:micro_us8k_performance_analysis}
	% \vspace{-8pt}
\end{figure}

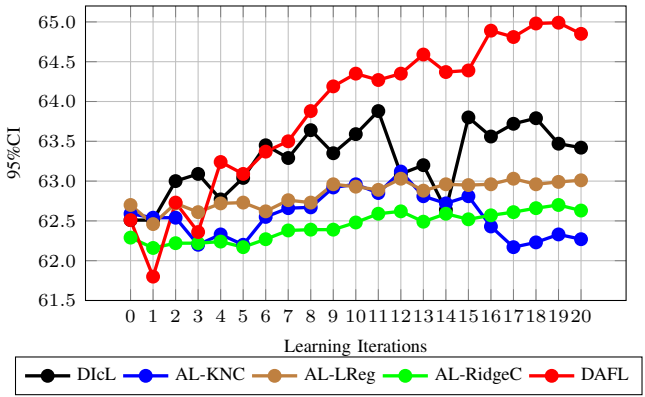
\begin{figure}[H]
	\centering
    \begin{tikzpicture}
        \begin{axis}[legend pos=north east,
            width = 0.99\columnwidth,
            height = 5.5cm,
            xlabel=Learning Iterations,
            grid=major,
            ylabel=\ac{95ci},
            xtick={0,1,2,3,4,5,6,7,8,9,10,11,12,13,14,15,16,17,18,19,20},
            % symbolic x coords={0,Loop1,Loop2,Loop3,Loop4,Loop5,Loop6,Loop7}, 
            ymin=61.5,ymax=65.2,
            ytick={61.5, 62, 62.5, 63, 63.5, 64, 64.5, 65},
            yticklabel style={/pgf/number format/.cd,fixed,fixed zerofill,precision=1},
            legend columns=5,
            ticklabel style = {font=\scriptsize},
            label style = {font=\scriptsize},
            legend style={at={(0.45,-0.19)}, anchor=north,legend cell align=left,font=\scriptsize} %
            ]
		\addplot[line width=1.25pt, color=black, mark=*] coordinates{
			(0, 62.51)(1, 62.51)(2, 63.00)(3, 63.09)(4, 62.77)(5, 63.04)
            (6, 63.45)(7, 63.29)(8, 63.64)(9, 63.35)(10, 63.59)
            (11, 63.88)(12, 63.09)(13, 63.20)(14, 62.64)(15, 63.80)
            (16, 63.56)(17, 63.72)(18, 63.79)(19, 63.47)(20, 63.42)
		};
        \addplot[line width=1.25pt, color=blue, mark=*] coordinates{
			(0, 62.59)(1, 62.54)(2, 62.54)(3, 62.20)(4, 62.33)(5, 62.20)
            (6, 62.55)(7, 62.66)(8, 62.67)(9, 62.92)(10, 62.96)
            (11, 62.85)(12, 63.12)(13, 62.81)(14, 62.72)(15, 62.81)
            (16, 62.43)(17, 62.17)(18, 62.23)(19, 62.33)(20, 62.27)
		};
        \addplot[line width=1.25pt, color=brown, mark=*] coordinates{
			(0, 62.70)(1, 62.46)(2, 62.72)(3, 62.61)(4, 62.72)(5, 62.73)
            (6, 62.62)(7, 62.76)(8, 62.73)(9, 62.96)(10, 62.93)
            (11, 62.89)(12, 63.03)(13, 62.88)(14, 62.96)(15, 62.95)
            (16, 62.96)(17, 63.03)(18, 62.96)(19, 62.99)(20, 63.01)
		};
        \addplot[line width=1.25pt, color=green, mark=*] coordinates{
			(0, 62.29)(1, 62.16)(2, 62.22)(3, 62.22)(4, 62.24)(5, 62.17)
            (6, 62.27)(7, 62.38)(8, 62.39)(9, 62.39)(10, 62.48)
            (11, 62.59)(12, 62.62)(13, 62.49)(14, 62.59)(15, 62.52)
            (16, 62.57)(17, 62.61)(18, 62.66)(19, 62.70)(20, 62.63)
		};
		\addplot[line width=1.25pt, color=red, mark=*] coordinates{
			(0, 62.51)(1, 61.80)(2, 62.73)(3, 62.36)(4, 63.24)(5, 63.09)
            (6, 63.37)(7, 63.50)(8, 63.88)(9, 64.19)(10, 64.35)
            (11, 64.27)(12, 64.35)(13, 64.59)(14, 64.37)(15, 64.39)
            (16, 64.89)(17, 64.81)(18, 64.98)(19, 64.99)(20, 64.85)
		};
		\legend{\ac{dicl}, \ac{alknc}, \ac{allgr}, \ac{alridgec}, \ac{dafl}}
		\end{axis}
	\end{tikzpicture}
    \vspace{-5pt}
    \caption{Edge-based \ac{dicl} vs \ac{dal} vs \ac{dafl} on \ac{iwbeat} dataset}
    \label{plot:micro_iwingbeat_performance_analysis}
	% \vspace{-8pt}
\end{figure}

\begin{figure}[H]
    \centering
    \vspace{-12pt}
    \includegraphics[width=1.0\linewidth]{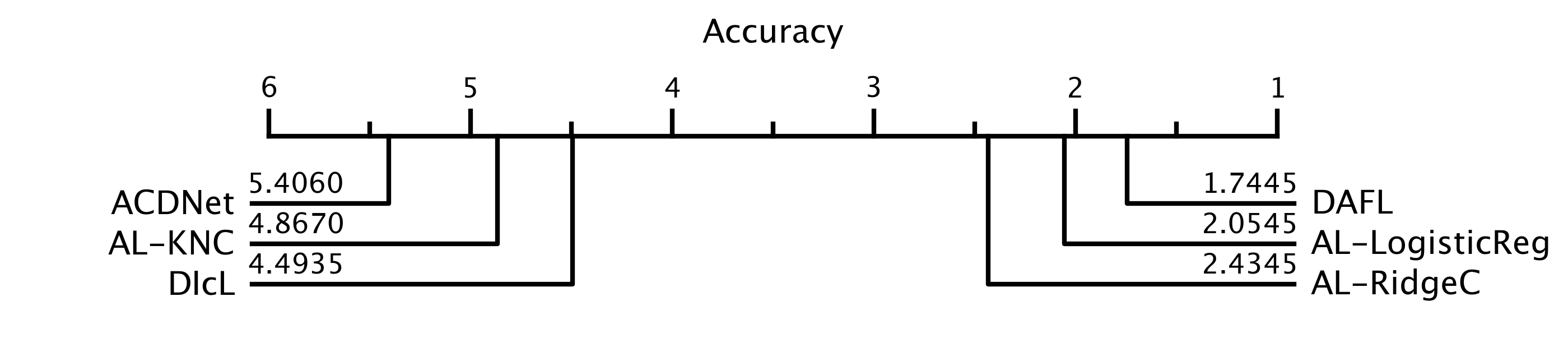}
    \vspace{-25pt}
    \caption{CD diagram showing statistical significance of the learning methods on the \ac{esc50} dataset.}
    \label{fig:micro_esc50_sig_diag}
    % \vspace{-8pt}
\end{figure}

\begin{figure}[H]
    \centering
    \vspace{-20pt}
    \includegraphics[width=1.0\linewidth]{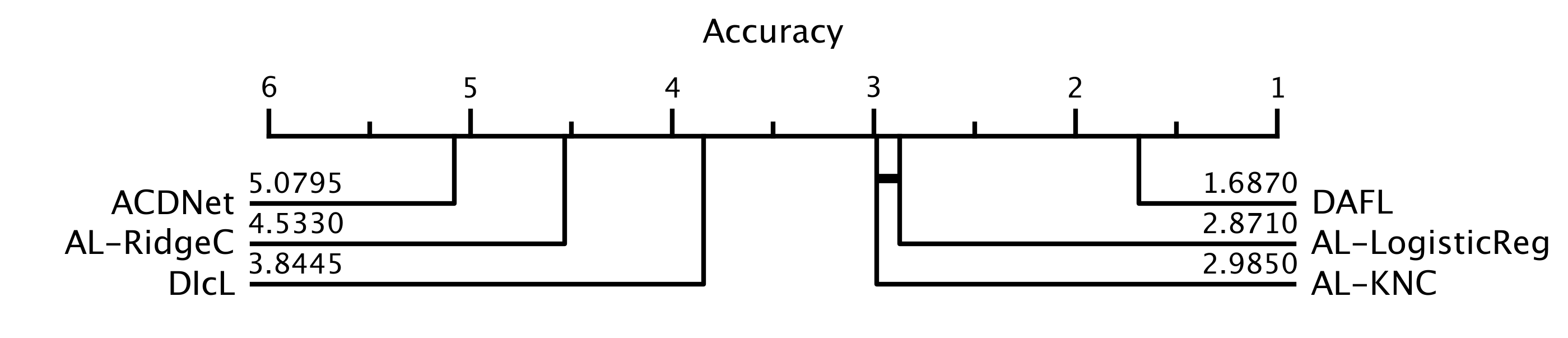}
    \vspace{-25pt}
    \caption{CD diagram showing statistical significance of the learning methods on the \ac{us8k} dataset.}
    \label{fig:micro_us8k_sig_diag}
    % \vspace{-8pt}
\end{figure}

\begin{figure}[H]
    \centering
    \vspace{-20pt}
    \includegraphics[width=1.0\linewidth]{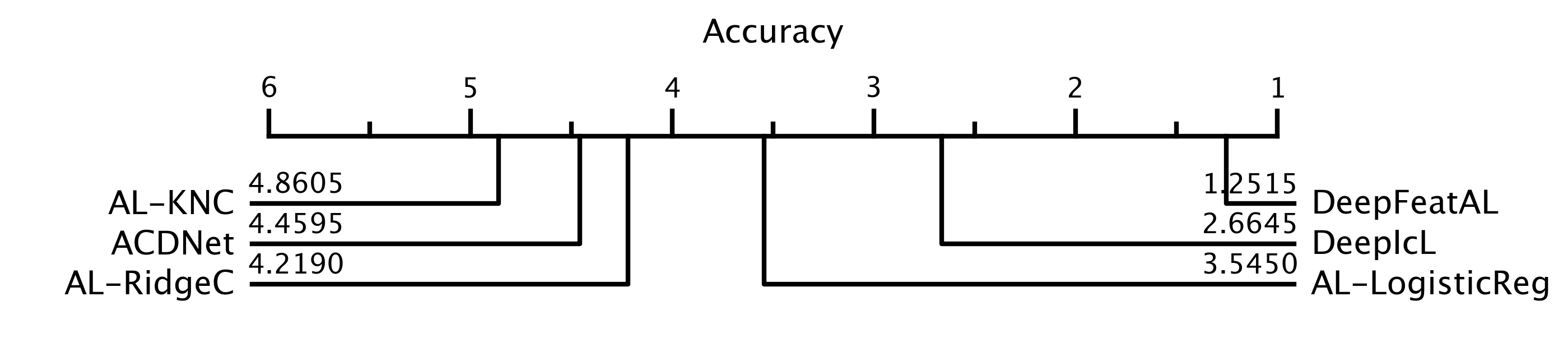}    \vspace{-25pt}
    \caption{CD diagram showing statistical significance of the learning methods on \ac{iwbeat} dataset.}
    \label{fig:micro_iwingbeat_sig_diag}
    % \vspace{-\baselineskip}
\end{figure}

The statistical significance of the increase in accuracy of the final models produced after completing all human annotation iterations on the \ac{esc50} dataset for \ac{dicl}, \ac{dal}, and \ac{dafl} is depicted in Figure~\ref{fig:micro_esc50_sig_diag}. According to the graph, all differences between models are statistically significant. It reveals that the \ac{dafl} model outperforms the other models and ranks first among them. 
The same holds for the experiments with the two other datasets (Figures~\ref{fig:micro_us8k_sig_diag} and \ref{fig:micro_iwingbeat_sig_diag}).
Overall, the results are qualitatively identical to those for the large network model: DAFL outperforms the conventional AL techniques in all cases. 


\section{Real-world Application}
\label{sec:al_btf}
To test the practical relevance of our method, we apply \ac{dafl} to data from a real-world conservation study.\footnote{Data provided by Lin Schwartzkopf and Slade Allen-Ankins (James Cook University, Townsville) and labelled by human experts under their supervision} The objective is to identify the calls of an endangered subspecies of the \acf{btf} ({\it Poephila cincta cincta}~\cite{dawe2005BTF}  found in Queensland, Australia)~\cite{dawe2005BTF} from audio  recorded in the wild with standard bio-acoustic recorders. 

The dataset consists of 49 files containing 98h of continuous audio recorded at 44.1kHz with 16-bit resolution ($\approx$2h per file). Calls were manually annotated by human experts and used to train and test an ACDnet model for the automated labelling of future recordings. A total of 2907 \ac{btf} calls were labelled.  Unlabelled data in the recordings consists of calls from various other species and environmental background sounds (BGS). 

Our review of the labelled data showed that the length of typical calls ranges from 0.6s to 0.672s. Consequently, we used sliding time windows of 0.6s length with a 0.1s offset for continuous recognition. Each sample is thus a vector $x_i \in \mathbb{R}^{12}$ (0.6s x 20,000Hz). We used zero padding at the end for windows that did not yield 12,000 data points. We augmented all \ac{btf} samples using \textit{adding noise, time-stretching, pitch-shifting, and time-shifting} to obtain 20,000 samples. We regarded a window as a positive sample if it overlapped with a BTF annotation by at least 50\% (see column Overlapping BTF Samples in Table~\ref{tab:btf_train_val_test_set_data_distribution}). Full details of the data sets are given in Table~\ref{tab:btf_train_val_test_set_data_distribution}.

\begin{table}[H]
    % \vspace{-5mm}
    \begin{center}
    \caption{Training set, validation set and test set details. The column Overlapping \ac{btf} Samples indicates samples having at least 50\% overlap with BTF annotations.}
    \resizebox{1.0\linewidth}{!}{%
    \begin{tabular}{| c | c | c | c | c | c | c | c |} 
    	\hline
      	Datasets & \thead{Recording\\ time (h)} & \#Files & \#Samples & \thead{Actual \\\ac{btf} Calls} & \thead{Overlapping \\\ac{btf} Samples} & \thead{BGS \\Samples} & \thead{Augmented \\\ac{btf} Samples}\\
      	\hline
      	Training set  & 80 & 40 & 410,349 & 2,602 & 10,349 & 4000,000 & 9,651\\
      	\hline
      	Validation set & 6 & 3 & 30,787 & 198 & 787 & 30,000 & -\\
        \hline
      	Test set  & 12 & 6 & 432,000 & 107 & 426 & 431,576 & -\\
      	\hline
    \end{tabular}}
    % \vspace{-1mm}
    \label{tab:btf_train_val_test_set_data_distribution}
  \end{center}
  % \vspace{-\baselineskip}
\end{table}

It is important to consider the exact nature of the real-world task as this has important implications for how performance must be measured. The objective is not to classify individual calls or to detect and/or count every individual call in the recordings. Rather, the real-world task is the detection of sites at which the target species is present. In a sense, the real task is the binary classification of sites, not sounds, according to the presence and absence of the target species. This site classification needs to be achieved with high accuracy and, importantly, with as little human intervention as possible. 

Since reliable site classification is required, verification by a human expert as a post-process cannot be avoided. This implies that a simple sliding time-window recognition as described above cannot  be used na\"ively: it will produce too many false positives,  necessitating too much manual checking. Even assuming an (unrealistically high) 95\% raw classification accuracy, na\"ive application would produce  more than 20,000 false positives per day (assuming equal rates of false positives and false negatives). It is impossible for a human expert to classify a 0.6s time window reliably in isolation. Rather, calls have to be listened to in a minimal temporal context (from experience, we assume 2 seconds). This means that 20,000 potential false positives amount to more than 11h human post-processing, so not much would be gained at all from such an approach  compared to direct listening.

To overcome this problem, we exploit the fact that calls generally occur in clusters if they are present at all. As we only need site classification, we can afford to miss \emph{most} calls at a site, as long as we do not miss \emph{all} calls. In theory, a single reliably identified call is enough to classify the site positively. This allows us to circumvent the above problem by strongly biasing our classifier towards false negatives and summarizing the recognition in longer segments for human checking. 

% \marginpar{\bf  Mohaimen, please update, amend this}
Based on an analysis of the call structures we decided to work with 5-second audio segments that were classified if they contained at least 4 recognized time slices. Biasing towards false negatives was induced using training class imbalance (see Table~\ref{tab:btf_train_val_test_set_data_distribution}).

Table~\ref{tab:btf_full_model_detection} and Table~\ref{tab:btf_micro_model_detection} present the results of applying ACDNet and Micro-ACDNet to the test data with this post-process. As each test file produced more than zero true positives (with the exception of Test File 5, which contains no calls), 100\% of the occupied sites would have been correctly detected by combing the automatic classification with human checking of just the positively classified segments (TP+FP). The effort for post-processing is determined by the number of these segments. ACD (Micro-ACDnet) produced a total of 21 TP and 8 FP (16 TP and 14 FP), respectively, for the whole test set. This amounts to ca.~2.5 minutes of human verification (in 5-second chunks) required for the whole 12-hour test data set, or equivalently less than 5 minutes per day---a very feasible amount. In practice, even less effort may be required, as the verification can be stopped as soon as the first positive sample has been verified with certainty. 

\begin{table}[htb!]
    % \vspace{-5mm}
    \caption{ACDNet prediction performance after post-processing}
    \begin{center}
    \resizebox{0.65\linewidth}{!}{%
    \begin{tabular}{| c | c | c | c | c | c | c |} 
    	\hline
      	\thead{Test \\File No.} & \thead{\#5s\\Segments} & \thead{\#\ac{btf} Call\\Segments} & TP & FP & FN & TN\\
      	\hline
      	1 & 1440 & 7 & 5 & 0 & 2 & 1433\\
      	\hline
      	2 & 1440 & 16 & 2 & 1 & 14 & 1423\\
        \hline
      	3 & 1440 & 12 & 3 & 1 & 9 & 1427\\
      	\hline
      	4 & 1440 & 23 & 6 & 2 & 17 & 1415\\
      	\hline
      	5 & 1440 & 0 & 0 & 1 & 0 & 1439\\
      	\hline
      	6 & 1440 & 13 & 5 & 3 & 8 & 1424\\
      	\hline
      	Total & 8,640 & 71 & 21 & 8 & 50 & 8,561\\
      	\hline
    \end{tabular}}
    % \vspace{-1mm}
    \label{tab:btf_full_model_detection}
  \end{center}
  % \vspace{-\baselineskip}
\end{table}

\begin{table}[htb!]
    \caption{Micro-ACDNet prediction performance after post-processing}
    \begin{center}
    \resizebox{0.65\linewidth}{!}{%
    \begin{tabular}{| c | c | c | c | c | c | c |} 
    	\hline
      	\thead{Test \\File No.} & \thead{\#5s\\Segments} & \thead{\#\ac{btf} Call\\Segments} & TP & FP & FN & TN\\
      	\hline
      	1 & 1440 & 7 & 4 & 0 & 3 & 1433\\
      	\hline
      	2 & 1440 & 16 & 2 & 3 & 14 & 1421\\
        \hline
      	3 & 1440 & 12 & 1 & 1 & 11 & 1427\\
      	\hline
      	4 & 1440 & 23 & 4 & 4 & 19 & 1413\\
      	\hline
      	5 & 1440 & 0 & 0 & 1 & 0 & 1439\\
      	\hline
      	6 & 1440 & 13 & 5 & 5 & 8 & 1422\\
      	\hline
      	Total & 8,640 & 71 & 16 & 14 & 55 & 8,555\\
      	\hline
    \end{tabular}}
    % \vspace{-1mm}
    \label{tab:btf_micro_model_detection}
  \end{center}
  % \vspace{-\baselineskip}
\end{table}

We now evaluate the relevance of the proposed \ac{dafl} schema in the context of this application. The full model reported above was trained on 80 hours of recordings that were  manually labelled by a human expert (Table~\ref{tab:btf_train_val_test_set_data_distribution}). We apply \ac{dafl} to reduce these labelling requirements. Initially, we use only 50\% of the training recordings to train a Micro-ACDnet. This equates to 40 hours of labelling effort. We then run 10 rounds of active learning. In each round, the algorithm queries the expert to label 1,000 samples selected by the active learning procedure. If we assume a 2-second context for the presentation of each sample, as above, this equates to ca.~33 minutes of labelling time per active learning iteration or 5.5 hours for all ten iterations. 

The performance of the \ac{dafl}-trained Micro-ACDnet model is fully comparable to the full-sized ACDnet model trained on the entire training set. Table~\ref{tab:btf_micro_al_full_retrain_details} shows the performance of the model after each round of active learning. The last two columns show how many samples from the 1,000 sample training batch presented by the algorithm in the specific training round are correctly classified before and after the re-training. To confirm that the increase in performance is indeed caused by the active learning and not just by further training epochs, we compare the active learning performance to the performance of the model trained on just the initial training set for the same additional number of epochs in every round (column ``retraining''). The model performing best in terms of overall precision is kept as the final trained model (Iteration 5). 

Table~\ref{tab:btf_micro_al_full_retrain_chosen_model} details the test performance of the final model for the six individual test files including post-processing as described above. We see that the number of true positives and false negatives is approximately equivalent to the full ACDnet model (Table~\ref{tab:btf_full_model_detection}). Likewise, the number of true positives plus false positives, which determines the total amount of human post-processing, differs only marginally (29 for the fully trained ACDnet versus 36 for the actively trained Micro-ACDnet).

Overall, the \ac{dafl} procedure allowed us to reduce the total labelling time required for training by 43\% from 80 hours to 45.5 hours while maintaining the same performance (in terms of human post-processing required for full detection accuracy). It did so while allowing us to simultaneously transition from a large model to a small one suitable for edge devices. This demonstrates the practical relevance of the proposed \ac{dafl}. 

\begin{table}[H]
    \caption{Micro-ACDNet performance for 10 rounds of active learning}
    \begin{center}
    \resizebox{1.0\linewidth}{!}{%
    \begin{tabular}{|c | c | c || c | c | c || c | c | c | c | c |} 
    	\hline
      	\thead{Learning \\Iteration} & \thead{\#5s \\Segments} & \thead{\#\ac{btf} \\Call \\Segments} & \multicolumn{3}{c||}{Retraining} & \multicolumn{5}{c|}{\ac{dafl}}\\
      	\cline{4-11}
      	& & & & & & & & & \multicolumn{2}{c|}{\thead{1,000 \\Annotated Samples}}\\
      	\cline{10-11}
      	& & & TP & FP & Precision & TP & FP & Precision & \thead{Known\\Before\\\gls{retrain}ing} & \thead{Known\\After\\\gls{retrain}ing}\\
      	\hline \hline
      	0 & 8,640 & 71 & 15 & 15 & 0.50 & 15 & 15 & 0.50 & - & -\\
        \hline
        1 & 8,640 & 71 & 19 & 22 & 0.46 & 21 & 21 & 0.50 & 646 & 730\\
        \hline
        2 & 8,640 & 71 & 21 & 38 & 0.36 & 19 & 18  & 0.51 & 655 & 767\\
        \hline
        3 & 8,640 & 71 & 21 & 47 & 0.31 & 17 & 20 & 0.46 & 709 & 747\\
        \hline
        4 & 8,640 & 71 & 19 & 23 & 0.45 & 21 & 24 & 0.47 & 787 & 799\\
        \hline
        {\bf 5} & {\bf 8,640} & {\bf 71 } & {\bf 22} & {\bf 34} & {\bf 0.39} & {\bf 19} & {\bf 17} & {\bf 0.53} & {\bf 775} & {\bf 818}\\
        \hline
        6 & 8,640 & 71 & 23 & 49 & 0.32 & 14 & 17 & 0.45 & 790 & 845\\
        \hline
        7 & 8,640 & 71 & 22 & 39 & 0.36 & 16 & 22 & 0.42 & 803 & 819\\
        \hline
        8 & 8,640 & 71 & 22 & 57 & 0.28 & 13 & 20 & 0.39 & 780 & 766\\
        \hline
        9 & 8,640 & 71 & 23 & 53 & 0.30 & 18 & 42 & 0.30 & 799 & 819\\
        \hline
        10 & 8,640 & 71 & 18 & 22 & 0.45 & 14 & 20 & 0.41 & 785 & 798\\
        \hline
    \end{tabular}}
    % \vspace{-1mm}
    \label{tab:btf_micro_al_full_retrain_details}
  \end{center}
  % \vspace{-\baselineskip}
\end{table}

\begin{table}[H]
    \caption{Best Micro-ACDNet model performance for each test file}
    \begin{center}
    \resizebox{0.65\linewidth}{!}{%
    \begin{tabular}{| c | c | c | c | c | c | c |} 
    	\hline
      	\thead{Test \\File No.} & \thead{\#5s\\Segments} & \thead{\#\ac{btf} Call\\Segments} & TP & FP & FN & TN\\
      	\hline
      	1 & 1440 & 7 & 4 & 1 & 3 & 1432\\
      	\hline
      	2 & 1440 & 16 & 3 & 4 & 13 & 1420\\
        \hline
      	3 & 1440 & 12 & 1 & 2 & 11 & 1426\\
      	\hline
      	4 & 1440 & 23 & 7 & 4 & 16 & 1412\\
      	\hline
       	5 & 1440 & 0 & 0 & 1 & 0 & 1439  \\
      	\hline
      	6 & 1440 & 13 & 4 & 5 & 9 & 1422\\
      	\hline
      	Total & 8,640 & 71 & 19 & 17 & 52 & 8,551\\
      	\hline
    \end{tabular}}
    % \vspace{-1mm}
    \label{tab:btf_micro_al_full_retrain_chosen_model}
  \end{center}
  % \vspace{-\baselineskip}
\end{table}
% \marginpar{\textcolor{red}{\bf Checked the numbers for file 4 and corrected the numbers.}}

\section{Conclusion}
Our study was designed to test the hypothesis that including feature extraction in the Active Learning loop provides performance benefits in (bio)acoustic classification.
Our experimental investigation on three widely used standard benchmark datasets (\ac{esc50}, \ac{us8k}, and \ac{iwbeat}) as well as on real-world data provides clear and statistically significant evidence to confirm this hypothesis. 

Integrating and \gls{ftuning} the feature extractor into the \ac{al} loop allows faster learning and thus enables us to either reach a set accuracy level with a smaller labelling budget or to reach higher accuracy with a fixed labelling budget. Given this, we find it worthwhile to explore the finer distinctions across different \ac{al} approaches, a work we want to pursue in the future.

The proposed method was tested with a large standard model and with a very small model suitable for edge-AI applications. It provided similar performance benefits in both cases. This should pave the way for the use of active learning in edge devices. 

In the future, we plan to investigate integrated edge-AI  architectures in which the edge devices autonomously collect new samples and send these ``back to base'' for expert labelling. The hope is that this will provide a useful approach for edge-AI devices that can continuously learn and improve their performance in the field. 

\section*{Acknowledgments}
This work was partly supported by the Australian Research Council under grant DE19010 0 045.

% \newpage
{\appendices
\section{Experimental Details}
\label{sec:al_experimenal_details}

% \marginpar{\bf I would think this belongs into an appendix or supplementary material.}
All experiments were conducted using Python version 3.7.4 and PyTorch 1.8.1. The full code is available at: \url{https://github.com/mohaimenz/deep_active_featl}.
\subsubsection{Datasets}
Experiments were carried out on three popular audio benchmark datasets: \ac{esc50}~\cite{piczak2015ESC}, \ac{us8k}~\cite{salamon2014Urbansound8k} and \ac{iwbeat}~\cite{chen2014FlyingInsect}. \ac{esc50} contains 2,000 samples, each of which is a 5-second audio recording sampled at 16kHz and 44.1kHz and distributed evenly among 50 balanced, separate classes (40 audio samples for each class). To achieve reproducible results, a presorted partition into 5 folds is also available for 5-fold cross-validation. The \ac{us8k} dataset contains 8,732 labelled audio clips (of 4-second length) of urban sounds recorded at 22.05kHz from 10 classes. The clips are grouped into 10 folds for cross-validating the results for fair comparison. The \ac{iwbeat} dataset includes 50,000 labelled audio clips (length of 1 second each) from 10 classes of insects, with 5,000 examples per class. We resampled all audio samples of all datasets to 20kHz.

\subsubsection{Splitting the Datasets for DeepIcL, Standard AL and DeepFeatAL}
We separate the datasets into two parts. One is used for training, validation, and testing, while the data from the second part are placed in an unlabeled data pool. We keep 50\%, 50\% and 35\% data instances from every class of the datasets for training, validation and testing.

We then create training sets, validation sets and test sets with 40\%, 20\% and 40\% of the reserved data (for training, validation and testing) of the \ac{esc50} dataset, 40\%, 20\% and 40\% of the \ac{us8k} dataset and 57\%, 28.5\% and 28.5\% of the \ac{iwbeat} dataset, respectively. 

Finally, we move the remaining data to the unlabeled data pool.

\subsubsection{Data Preprocessing}
For \ac{esc50} and \ac{us8k}, we follow the procedure described in \cite{mohaimen2021ACDNet}. For the \ac{iwbeat} dataset, we use input samples of length 20,000, or 1s audio at 20kHz since the recordings are only 1s long. We use the data augmentation techniques described in \citet{mohaimen2021ACDNet} and~\citet{mohaimen2021pruningvsxnor} to augment the training sets of the \ac{esc50}, \ac{us8k} and \ac{iwbeat} datasets. However, we do not employ the mixup of two classes for the training set augmentation and 10-crops of test data proposed in those articles because our goal is to improve model learning via active learning rather than those strategies.

\subsubsection{Hyperparameter Settings for Initial Training}
ACDNet is trained for 600 epochs with a learning rate scheduler {0.3, 0.6, 0.9} with the same learning settings described in \citet{mohaimen2021ACDNet}. At the end of training and validation, we use the best model for testing, \ac{dicl}, \ac{dal} and \ac{dafl}.

\subsubsection{Learning Settings for DeepIcl, Standard AL and DeepFeatAL}
We use seven, fifteen, and twenty iterations of human annotation (simulated) on the \ac{esc50}, \ac{us8k}, and \ac{iwbeat} datasets, respectively. In each iteration, the simulated annotation system is requested to provide the labels of 100 specific unlabeled samples (determined by the acquisition function) for the first two datasets and 500 unlabeled samples for the third dataset. In each iteration of the simulated labelling, we \gls{ftune} the model for 100 epochs for each dataset during \ac{dicl} and \ac{dafl}; however, we add the same number of old training data to the newly labelled data to prevent the network from forgetting its prior knowledge. During this stage of learning with new data, we employ a significantly lower learning rate (i.e., 0.001), a new learning rate scheduler of [15, 60, 90] without any warm-ups and a batch size of 16.

\section{Tables Representing Experimental Results}
\label{appendix_addtional_tables}
\begin{table}[H]
    \caption{ACDNet in different training settings for incremental Learning. 
    % The first row indicates the performance of the model before incremental learning.
    } 
    \begin{center}
    \resizebox{0.65\linewidth}{!}{%
    \begin{tabular}{| c | c | c | c |} 
    	\hline
      	&\multicolumn{3}{c|}{Prediction Accuracy (\%)}\\
      	\cline{2-4}
      	\thead{Learning \\Loops}& \Gls{nfreeze} & \Gls{ffreeze} & \Gls{sfreeze}\\
      	\hline
      	0 & 55.75 & 55.75 & 55.75\\
      	\hline
      	1 & 59.50 & 58.00 & 59.50\\
        \hline
      	2 & 61.25 & 59.25 & 60.75\\
      	\hline
        3 & 61.75 & 59.75 & 61.25\\
      	\hline
      	4 & 64.00 & 62.00 & 63.00\\
      	\hline
      	5 & 63.50 & 62.50 & 62.00\\
      	\hline
      	6 & 64.75 & 62.75 & 63.75\\
      	\hline
      	7 & \textbf{65.50} & 63.50 & 64.00\\
      	\hline
    \end{tabular}}
    % \vspace{-1mm}
    \label{tab:al_acdnet_freeze_nofreeze_esc50}
  \end{center}
  % \vspace{-8mm}
\end{table}

\begin{table*}
    \caption{\ac{dicl} vs \ac{dal} vs \ac{dafl} using ACDNet on \ac{esc50} dataset}
    \begin{center}
    \resizebox{0.6\linewidth}{!}{%
    \begin{tabular}{| c | c | c | c | c| c|} 
    	\hline
      	&\multicolumn{5}{c|}{Accuracy(\ac{95ci})}\\
      	\cline{2-6}
        &&\multicolumn{3}{c|}{Other \acf{al}}&\\
        \cline{3-5}
      	\thead{Learning \\Loops}& \ac{dicl} & \ac{alknc} & \ac{allgr} & \ac{alridgec} & \ac{dafl}(proposed)\\
      	\hline
      	0 & 55.81 $\pm$ 0.15 & 54.02 $\pm$ 0.15 & 58.78 $\pm$ 0.15  & 57.79 $\pm$ 0.16 & 55.81 $\pm$ 0.16\\
      	\hline
      	1 & 59.56 $\pm$ 0.15 & 56.83 $\pm$ 0.14 & 60.76 $\pm$ 0.16  & 59.54 $\pm$ 0.14 & 63.76 $\pm$ 0.14\\
        \hline
      	2 & 61.23 $\pm$ 0.15 & 61.19 $\pm$ 0.15 & 63.19 $\pm$ 0.15  & 62.58 $\pm$ 0.15 & 64.15 $\pm$ 0.15\\
      	\hline
      	3 & 61.71 $\pm$ 0.15 & 62.73 $\pm$ 0.15 & 63.81 $\pm$ 0.16  & 65.19 $\pm$ 0.14 & 65.00 $\pm$ 0.16\\
      	\hline
      	4 & 63.94 $\pm$ 0.15 & 63.19 $\pm$ 0.14 & 64.04 $\pm$ 0.15  & 67.00 $\pm$ 0.15 & 65.94 $\pm$ 0.14\\
      	\hline
      	5 & 63.38 $\pm$ 0.15 & 63.22 $\pm$ 0.15 & 65.46 $\pm$ 0.15  & 67.10 $\pm$ 0.15 & 67.28 $\pm$ 0.15\\
      	\hline
      	6 & 64.69 $\pm$ 0.15 & 64.86 $\pm$ 0.14 & 64.50 $\pm$ 0.15  & 67.73 $\pm$ 0.15 & 68.74 $\pm$ 0.14\\
      	\hline
      	7 & 65.43 $\pm$ 0.15 & 65.18 $\pm$ 0.15 & 66.77 $\pm$ 0.15  & 67.94 $\pm$ 0.14 & 70.02 $\pm$ 0.12\\
      	\hline
    \end{tabular}}
    \label{tab:al_icl_vs_dal_dafl_esc50}
  \end{center}
  % \vspace{-8mm}
\end{table*}

\begin{table*}
    \caption{\ac{dicl} vs \ac{dal} vs \ac{dafl} on \ac{us8k} dataset}
    \begin{center}
    \resizebox{0.6\linewidth}{!}{%
    \begin{tabular}{| c | c | c | c | c| c|} 
    	\hline
      	&\multicolumn{5}{c|}{Accuracy(\ac{95ci})}\\
      	\cline{2-6}
        &&\multicolumn{3}{c|}{Other \acf{al}}&\\
        \cline{3-5}
      	\thead{Learning \\Loops}& \ac{dicl} & \ac{alknc} & \ac{allgr} & \ac{alridgec} & \ac{dafl} (proposed)\\
      	\hline
      	0 & 86.97 $\pm$ 0.06 & 87.24 $\pm$ 0.05 & 87.55 $\pm$ 0.05 & 87.00 $\pm$ 0.05 & 86.97 $\pm$ 0.06\\
      	\hline
      	1 & 87.12 $\pm$ 0.05 & 87.60 $\pm$ 0.05 & 87.48 $\pm$ 0.05 & 86.80 $\pm$ 0.05 & 85.93 $\pm$ 0.05\\
        \hline
      	2 & 86.94 $\pm$ 0.04 & 87.51 $\pm$ 0.05 & 88.25 $\pm$ 0.04 & 87.09 $\pm$ 0.05 & 86.82 $\pm$ 0.05\\
      	\hline
      	3 & 87.47 $\pm$ 0.06 & 87.49 $\pm$ 0.05 & 88.08 $\pm$ 0.04 & 87.60 $\pm$ 0.04 & 87.15 $\pm$ 0.05\\
      	\hline
      	4 & 87.46 $\pm$ 0.05 & 87.66 $\pm$ 0.05 & 88.00 $\pm$ 0.05 & 87.42 $\pm$ 0.05 & 87.09 $\pm$ 0.05\\
      	\hline
      	5 & 87.98 $\pm$ 0.04 & 87.65 $\pm$ 0.04 & 87.96 $\pm$ 0.05 & 87.27 $\pm$ 0.06 & 88.59 $\pm$ 0.06\\
      	\hline
      	6 & 87.80 $\pm$ 0.05 & 88.05 $\pm$ 0.05 & 87.96 $\pm$ 0.05 & 87.38 $\pm$ 0.05 & 88.20 $\pm$ 0.05\\
      	\hline
      	7 & 87.78 $\pm$ 0.06 & 88.14 $\pm$ 0.05 & 87.79 $\pm$ 0.05 & 87.47 $\pm$ 0.04 & 88.94 $\pm$ 0.06\\
      	\hline
      	8 & 87.94 $\pm$ 0.05 & 88.12 $\pm$ 0.04 & 88.08 $\pm$ 0.04 & 87.54 $\pm$ 0.05 & 88.62 $\pm$ 0.05\\
      	\hline
      	9 & 86.91 $\pm$ 0.05 & 88.04 $\pm$ 0.05 & 88.16 $\pm$ 0.05 & 87.49 $\pm$ 0.05 & 88.60 $\pm$ 0.05\\
      	\hline
      	10 & 87.67 $\pm$ 0.05 & 87.93 $\pm$ 0.05 & 88.25 $\pm$ 0.04 & 87.72 $\pm$ 0.04 & 88.70 $\pm$ 0.05\\
      	\hline
      	11 & 88.19 $\pm$ 0.06 & 88.05 $\pm$ 0.05 & 88.18 $\pm$ 0.05 & 87.70 $\pm$ 0.06 & 88.75 $\pm$ 0.06\\
      	\hline
      	12 & 87.67 $\pm$ 0.05 & 87.91 $\pm$ 0.05 & 88.21 $\pm$ 0.05 & 87.73 $\pm$ 0.05 & 88.80 $\pm$ 0.05\\
      	\hline
      	13 & 87.00 $\pm$ 0.04 & 87.80 $\pm$ 0.05 & 88.48 $\pm$ 0.06 & 87.80 $\pm$ 0.05 & 89.01 $\pm$ 0.06\\
      	\hline
      	14 & 86.71 $\pm$ 0.06 & 88.18 $\pm$ 0.05 & 88.17 $\pm$ 0.05 & 87.78 $\pm$ 0.05 & 89.20 $\pm$ 0.06\\
      	\hline
      	15 & 87.38 $\pm$ 0.04 & 88.48 $\pm$ 0.05 & 88.31 $\pm$ 0.05 & 87.82 $\pm$ 0.05 & 89.45 $\pm$ 0.05\\
      	\hline
    \end{tabular}}
    % \vspace{-1mm}
    \label{tab:al_icl_vs_dal_vs_dafl_us8k}
  \end{center}
  % \vspace{-8mm}
\end{table*}

\begin{table*}
    \caption{\ac{dicl} vs \ac{dal} vs \ac{dafl} on \ac{iwbeat} dataset}
    \begin{center}
    \resizebox{0.6\linewidth}{!}{%
    \begin{tabular}{| c | c | c | c | c| c|} 
    	\hline
      	&\multicolumn{5}{c|}{Accuracy(\ac{95ci})}\\
      	\cline{2-6}
        &&\multicolumn{3}{c|}{Other \acf{al}}&\\
        \cline{3-5}
      	\thead{Learning \\Loops}& \ac{dicl} & \ac{alknc} & \ac{allgr} & \ac{alridgec} & \ac{dafl}(proposed)\\
      	\hline
      	0 & 66.08 $\pm$ 0.04 & 66.31 $\pm$ 0.04 & 66.11 $\pm$ 0.04 & 65.82 $\pm$ 0.04 & 66.08 $\pm$ 0.04\\
      	\hline
      	1 & 66.94 $\pm$ 0.04 & 66.31 $\pm$ 0.04 & 66.53 $\pm$ 0.04 & 65.79 $\pm$ 0.04 & 66.59 $\pm$ 0.04\\
        \hline
      	2 & 66.50 $\pm$ 0.04 & 66.47 $\pm$ 0.04 & 66.57 $\pm$ 0.04 & 65.95 $\pm$ 0.04 & 66.99 $\pm$ 0.04\\
      	\hline
      	3 & 66.21 $\pm$ 0.04 & 66.61 $\pm$ 0.04 & 66.53 $\pm$ 0.04 & 66.02 $\pm$ 0.04 & 67.57 $\pm$ 0.04\\
      	\hline
      	4 & 66.63 $\pm$ 0.04 & 66.44 $\pm$ 0.04 & 66.64 $\pm$ 0.04 & 65.89 $\pm$ 0.04 & 67.13 $\pm$ 0.04\\
      	\hline
      	5 & 67.19 $\pm$ 0.04 & 66.22 $\pm$ 0.04 & 66.61 $\pm$ 0.04 & 65.87 $\pm$ 0.04 & 67.40 $\pm$ 0.04\\
      	\hline
      	6 & 67.30 $\pm$ 0.04 & 66.17 $\pm$ 0.04 & 66.68 $\pm$ 0.04 & 66.03 $\pm$ 0.04 & 67.76 $\pm$ 0.04\\
      	\hline
      	7 & 67.79 $\pm$ 0.04 & 66.19 $\pm$ 0.04 & 66.79 $\pm$ 0.04 & 66.01 $\pm$ 0.04 & 67.66 $\pm$ 0.04\\
      	\hline
      	8 & 67.11 $\pm$ 0.04 & 66.29 $\pm$ 0.04 & 66.91 $\pm$ 0.04 & 66.14 $\pm$ 0.04 & 68.10 $\pm$ 0.04\\
      	\hline
      	9 & 67.32 $\pm$ 0.04 & 66.28 $\pm$ 0.04 & 66.95 $\pm$ 0.04 & 66.04 $\pm$ 0.04 & 68.03 $\pm$ 0.04\\
      	\hline
      	10 & 67.20 $\pm$ 0.04 & 66.14 $\pm$ 0.04 & 67.02 $\pm$ 0.04 & 66.16 $\pm$ 0.04 & 68.86 $\pm$ 0.04\\
      	\hline
      	11 & 67.97 $\pm$ 0.04 & 66.24 $\pm$ 0.04 & 67.15 $\pm$ 0.04 & 66.12 $\pm$ 0.04 & 68.95 $\pm$ 0.04\\
      	\hline
      	12 & 67.07 $\pm$ 0.04 & 66.15 $\pm$ 0.04 & 67.05 $\pm$ 0.04 & 66.36 $\pm$ 0.04 & 69.90 $\pm$ 0.04\\
      	\hline
      	13 & 67.27 $\pm$ 0.04 & 66.22 $\pm$ 0.04 & 67.10 $\pm$ 0.04 & 66.32 $\pm$ 0.04 & 69.05 $\pm$ 0.04\\
      	\hline
      	14 & 66.66 $\pm$ 0.04 & 66.35 $\pm$ 0.04 & 67.10 $\pm$ 0.04 & 66.20 $\pm$ 0.04 & 69.29 $\pm$ 0.04\\
      	\hline
      	15 & 67.54 $\pm$ 0.04 & 66.35 $\pm$ 0.04 & 67.03 $\pm$ 0.04 & 66.14 $\pm$ 0.04 & 69.32 $\pm$ 0.04\\
      	\hline
      	16 & 67.50 $\pm$ 0.04 & 66.43 $\pm$ 0.04 & 67.13 $\pm$ 0.04 & 66.25 $\pm$ 0.04 & 69.76 $\pm$ 0.04\\
      	\hline
      	17 & 67.65 $\pm$ 0.04 & 66.10 $\pm$ 0.04 & 67.17 $\pm$ 0.04 & 66.31 $\pm$ 0.04 & 70.12 $\pm$ 0.04\\
      	\hline
      	18 & 67.62 $\pm$ 0.04 & 66.11 $\pm$ 0.04 & 67.17 $\pm$ 0.04 & 66.37 $\pm$ 0.04 & 70.04 $\pm$ 0.04\\
      	\hline
      	19 & 68.27 $\pm$ 0.04 & 66.03 $\pm$ 0.04 & 67.11 $\pm$ 0.04 & 66.30 $\pm$ 0.04 & 70.18 $\pm$ 0.04\\
      	\hline
      	20 & 68.16 $\pm$ 0.04 & 65.87 $\pm$ 0.04 & 67.17 $\pm$ 0.04 & 66.28 $\pm$ 0.04 & 70.13 $\pm$ 0.04\\
      	\hline
    \end{tabular}}
    % \vspace{-1mm}
    \label{tab:al_icl_vs_dal_vs_dafl_iwingbeat}
  \end{center}
  % \vspace{-8mm}
\end{table*}

\begin{table*}
    \caption{Edge-based \ac{dicl} vs \ac{dal} vs \ac{dafl} on \ac{esc50} dataset}
    \begin{center}
    \resizebox{0.6\linewidth}{!}{%
    \begin{tabular}{| c | c | c | c | c| c|} 
    	\hline
      	&\multicolumn{5}{c|}{Accuracy(\ac{95ci})}\\
      	\cline{2-6}
        &&\multicolumn{3}{c|}{Other \acf{al}}&\\
        \cline{3-5}
      	\thead{Learning \\Loops}& \ac{dicl} & \ac{alknc} & \ac{allgr} & \ac{alridgec} & \ac{dafl}(proposed)\\
      	\hline
      	0 & 55.36 $\pm$ 0.14 & 47.89 $\pm$ 0.15 & 52.67 $\pm$ 0.15  & 52.46 $\pm$ 0.16 & 55.36 $\pm$ 0.14\\
      	\hline
      	1 & 55.99 $\pm$ 0.15 & 48.89 $\pm$ 0.14 & 56.08 $\pm$ 0.16  & 53.89 $\pm$ 0.14 & 58.31 $\pm$ 0.15\\
        \hline
      	2 & 55.72 $\pm$ 0.15 & 53.80 $\pm$ 0.15 & 56.47 $\pm$ 0.15  & 58.09 $\pm$ 0.15 & 60.05 $\pm$ 0.15\\
      	\hline
      	3 & 55.24 $\pm$ 0.15 & 55.35 $\pm$ 0.15 & 60.25 $\pm$ 0.16  & 61.53 $\pm$ 0.14 & 60.76 $\pm$ 0.15\\
      	\hline
      	4 & 56.25 $\pm$ 0.15 & 58.83 $\pm$ 0.14 & 60.25 $\pm$ 0.15  & 61.16 $\pm$ 0.15 & 63.61 $\pm$ 0.15\\
      	\hline
      	5 & 54.94 $\pm$ 0.15 & 57.37 $\pm$ 0.15 & 62.76 $\pm$ 0.15  & 61.20 $\pm$ 0.15 & 63.25 $\pm$ 0.15\\
      	\hline
      	6 & 57.37 $\pm$ 0.15 & 57.92 $\pm$ 0.14 & 62.70 $\pm$ 0.15  & 60.70 $\pm$ 0.15 & 64.01 $\pm$ 0.15\\
      	\hline
      	7 & \textbf{57.70 $\pm$ 0.15} & 56.67 $\pm$ 0.15 & 63.32 $\pm$ 0.15  & 62.38 $\pm$ 0.14 & 65.20 $\pm$ 0.15\\
      	\hline
    \end{tabular}}
    \vspace{-1mm}
    \label{tab:al_micro_icl_vs_dal_vs_dafl_esc50}
  \end{center}
  % \vspace{-8mm}
\end{table*}

\begin{table*}
    \caption{Edge-based \ac{dicl} vs \ac{dal} vs \ac{dafl} on \ac{us8k} dataset}
    \begin{center}
    \resizebox{0.6\linewidth}{!}{%
    \begin{tabular}{| c | c | c | c | c| c|} 
    	\hline
      	&\multicolumn{5}{c|}{Accuracy(\ac{95ci})}\\
      	\cline{2-6}
        &&\multicolumn{3}{c|}{Other \acf{al}}&\\
        \cline{3-5}
      	\thead{Learning \\Loops}& \ac{dicl} & \ac{alknc} & \ac{allgr} & \ac{alridgec} & \ac{dafl}(proposed)\\
      	\hline
      	0 & 81.71 $\pm$ 0.06 & 81.77 $\pm$ 0.06 & 82.79 $\pm$ 0.06 & 81.84 $\pm$ 0.06 & 81.73 $\pm$ 0.05\\
      	\hline
      	1 & 81.73 $\pm$ 0.06 & 82.06 $\pm$ 0.06 & 82.70 $\pm$ 0.06 & 81.71 $\pm$ 0.06 & 81.97 $\pm$ 0.05\\
        \hline
      	2 & 81.41 $\pm$ 0.06 & 82.16 $\pm$ 0.06 & 82.70 $\pm$ 0.06 & 82.02 $\pm$ 0.05 & 81.95 $\pm$ 0.05\\
      	\hline
      	3 & 82.57 $\pm$ 0.06 & 82.36 $\pm$ 0.06 & 82.92 $\pm$ 0.06 & 82.26 $\pm$ 0.06 & 82.88 $\pm$ 0.06\\
      	\hline
      	4 & 82.70 $\pm$ 0.06 & 82.44 $\pm$ 0.06 & 82.85 $\pm$ 0.06 & 81.92 $\pm$ 0.06 & 83.03 $\pm$ 0.05\\
      	\hline
      	5 & 83.13 $\pm$ 0.06 & 82.20 $\pm$ 0.06 & 83.09 $\pm$ 0.06 & 82.04 $\pm$ 0.06 & 83.56 $\pm$ 0.05\\
      	\hline
      	6 & 82.34 $\pm$ 0.06 & 82.35 $\pm$ 0.06 & 83.11 $\pm$ 0.06 & 81.98 $\pm$ 0.06 & 83.56 $\pm$ 0.05\\
      	\hline
      	7 & 82.98 $\pm$ 0.06 & 82.85 $\pm$ 0.06 & 82.79 $\pm$ 0.06 & 82.12 $\pm$ 0.06 & 83.15 $\pm$ 0.05\\
      	\hline
      	8 & 82.76 $\pm$ 0.06 & 82.80 $\pm$ 0.06 & 83.11 $\pm$ 0.06 & 81.96 $\pm$ 0.06 & 83.13 $\pm$ 0.05\\
      	\hline
      	9 & 82.10 $\pm$ 0.06 & 82.77 $\pm$ 0.06 & 83.18 $\pm$ 0.06 & 82.01 $\pm$ 0.06 & 83.46 $\pm$ 0.05\\
      	\hline
      	10 & 82.36 $\pm$ 0.06 & 82.66 $\pm$ 0.06 & 83.10 $\pm$ 0.06 & 82.25 $\pm$ 0.06 & 83.41 $\pm$ 0.05\\
      	\hline
      	11 & 82.33 $\pm$ 0.06 & 83.02 $\pm$ 0.06 & 83.17 $\pm$ 0.06 & 82.16 $\pm$ 0.06 & 83.48 $\pm$ 0.05\\
      	\hline
      	12 & 82.51 $\pm$ 0.06 & 83.30 $\pm$ 0.06 & 83.38 $\pm$ 0.06 & 82.01 $\pm$ 0.06 & 84.35 $\pm$ 0.05\\
      	\hline
      	13 & 82.80 $\pm$ 0.06 & 83.24 $\pm$ 0.06 & 83.19 $\pm$ 0.06 & 82.15 $\pm$ 0.06 & 84.03 $\pm$ 0.05\\
      	\hline
      	14 & 82.64 $\pm$ 0.06 & 82.97 $\pm$ 0.06 & 83.20 $\pm$ 0.06 & 81.89 $\pm$ 0.06 & 83.99 $\pm$ 0.05\\
      	\hline
      	15 & 82.63 $\pm$ 0.06 & 83.15 $\pm$ 0.06 & 83.27 $\pm$ 0.06 & 82.16 $\pm$ 0.06 & 84.12 $\pm$ 0.05\\
      	\hline
    \end{tabular}}
    % \vspace{-1mm}
    \label{tab:al_micro_icl_vs_dal_vs_dafl_us8k}
  \end{center}
  % \vspace{-8mm}
\end{table*}

\begin{table*}
    \caption{Edge-based \ac{dicl} vs \ac{dal} vs \ac{dafl} on \ac{iwbeat} dataset}
    \begin{center}
    \resizebox{0.6\linewidth}{!}{%
    \begin{tabular}{| c | c | c | c | c| c|} 
    	\hline
      	&\multicolumn{5}{c|}{Accuracy(\ac{95ci})}\\
      	\cline{2-6}
        &&\multicolumn{3}{c|}{Other \acf{al}}&\\
        \cline{3-5}
      	\thead{Learning \\Loops}& \ac{dicl} & \ac{alknc} & \ac{allgr} & \ac{alridgec} & \ac{dafl}(proposed)\\
      	\hline
      	0 & 62.51 $\pm$ 0.04 & 62.59 $\pm$ 0.04 & 62.70 $\pm$ 0.04 & 62.29 $\pm$ 0.04 & 62.51 $\pm$ 0.04\\
      	\hline
      	1 & 62.51 $\pm$ 0.05 & 62.54 $\pm$ 0.04 & 62.46 $\pm$ 0.04 & 62.16 $\pm$ 0.04 & 61.80 $\pm$ 0.04\\
        \hline
      	2 & 63.00 $\pm$ 0.05 & 62.54 $\pm$ 0.04 & 62.72 $\pm$ 0.04 & 62.22 $\pm$ 0.04 & 62.73 $\pm$ 0.04\\
      	\hline
      	3 & 63.09 $\pm$ 0.04 & 62.20 $\pm$ 0.04 & 62.61 $\pm$ 0.04 & 62.22 $\pm$ 0.04 & 62.36 $\pm$ 0.04\\
      	\hline
      	4 & 62.77 $\pm$ 0.04 & 62.33 $\pm$ 0.04 & 62.72 $\pm$ 0.04 & 62.24 $\pm$ 0.04 & 63.24 $\pm$ 0.04\\
      	\hline
      	5 & 63.04 $\pm$ 0.04 & 62.20 $\pm$ 0.04 & 62.73 $\pm$ 0.04 & 62.17 $\pm$ 0.04 & 63.09 $\pm$ 0.04\\
      	\hline
      	6 & 63.45 $\pm$ 0.05 & 62.55 $\pm$ 0.04 & 62.62 $\pm$ 0.04 & 62.27 $\pm$ 0.04 & 63.37 $\pm$ 0.04\\
      	\hline
      	7 & 63.29 $\pm$ 0.04 & 62.66 $\pm$ 0.04 & 62.76 $\pm$ 0.04 & 62.38 $\pm$ 0.04 & 63.50 $\pm$ 0.04\\
      	\hline
      	8 & 63.64 $\pm$ 0.04 & 62.67 $\pm$ 0.04 & 62.73 $\pm$ 0.04 & 62.39 $\pm$ 0.04 & 63.88 $\pm$ 0.04\\
      	\hline
      	9 & 63.35 $\pm$ 0.04 & 62.92 $\pm$ 0.04 & 62.96 $\pm$ 0.04 & 62.39 $\pm$ 0.04 & 64.19 $\pm$ 0.04\\
      	\hline
      	10 & 63.59 $\pm$ 0.04 & 62.96 $\pm$ 0.04 & 62.93 $\pm$ 0.04 & 62.48 $\pm$ 0.04 & 64.35 $\pm$ 0.04\\
      	\hline
      	11 & 63.88 $\pm$ 0.04 & 62.85 $\pm$ 0.04 & 62.89 $\pm$ 0.04 & 62.59 $\pm$ 0.04 & 64.27 $\pm$ 0.04\\
      	\hline
      	12 & 63.09 $\pm$ 0.05 & 63.12 $\pm$ 0.04 & 63.03 $\pm$ 0.04 & 62.62 $\pm$ 0.04 & 64.35 $\pm$ 0.04\\
      	\hline
      	13 & 63.20 $\pm$ 0.05 & 62.81 $\pm$ 0.04 & 62.88 $\pm$ 0.04 & 62.49 $\pm$ 0.04 & 64.59 $\pm$ 0.04\\
      	\hline
      	14 & 62.64 $\pm$ 0.05 & 62.72 $\pm$ 0.04 & 62.96 $\pm$ 0.04 & 62.59 $\pm$ 0.04 & 64.37 $\pm$ 0.04\\
      	\hline
      	15 & 63.80 $\pm$ 0.04 & 62.81 $\pm$ 0.04 & 62.95 $\pm$ 0.04 & 62.52 $\pm$ 0.04 & 64.39 $\pm$ 0.04\\
      	\hline
      	16 & 63.56 $\pm$ 0.04 & 62.43 $\pm$ 0.04 & 62.96 $\pm$ 0.04 & 62.57 $\pm$ 0.04 & 64.89 $\pm$ 0.04\\
      	\hline
      	17 & 63.72 $\pm$ 0.04 & 62.17 $\pm$ 0.04 & 63.03 $\pm$ 0.04 & 62.61 $\pm$ 0.04 & 64.81 $\pm$ 0.04\\
      	\hline
      	18 & 63.79 $\pm$ 0.04 & 62.23 $\pm$ 0.04 & 62.96 $\pm$ 0.04 & 62.66 $\pm$ 0.04 & 64.98 $\pm$ 0.04\\
      	\hline
      	19 & 63.47 $\pm$ 0.04 & 62.33 $\pm$ 0.04 & 62.99 $\pm$ 0.04 & 62.70 $\pm$ 0.04 & 64.99 $\pm$ 0.04\\
      	\hline
      	20 & 63.42 $\pm$ 0.04 & 62.27 $\pm$ 0.04 & 63.01 $\pm$ 0.04 & 62.63 $\pm$ 0.04 & 64.85 $\pm$ 0.04\\
      	\hline
    \end{tabular}}
    % \vspace{-1mm}
    \label{tab:al_micro_icl_vs_dal_vs_dafl_iwingbeat}
  \end{center}
  % \vspace{-8mm}
\end{table*}
}

\newpage
\scriptsize
\bibliographystyle{IEEEtranN}
\bibliography{ReferenceMaster}
% \end{thebibliography}
\normalsize

% \newpage

\section*{Author Biography}
\vspace{-20pt}
\begin{IEEEbiography}[{\includegraphics[width=1in,height=1.25in,clip,keepaspectratio]{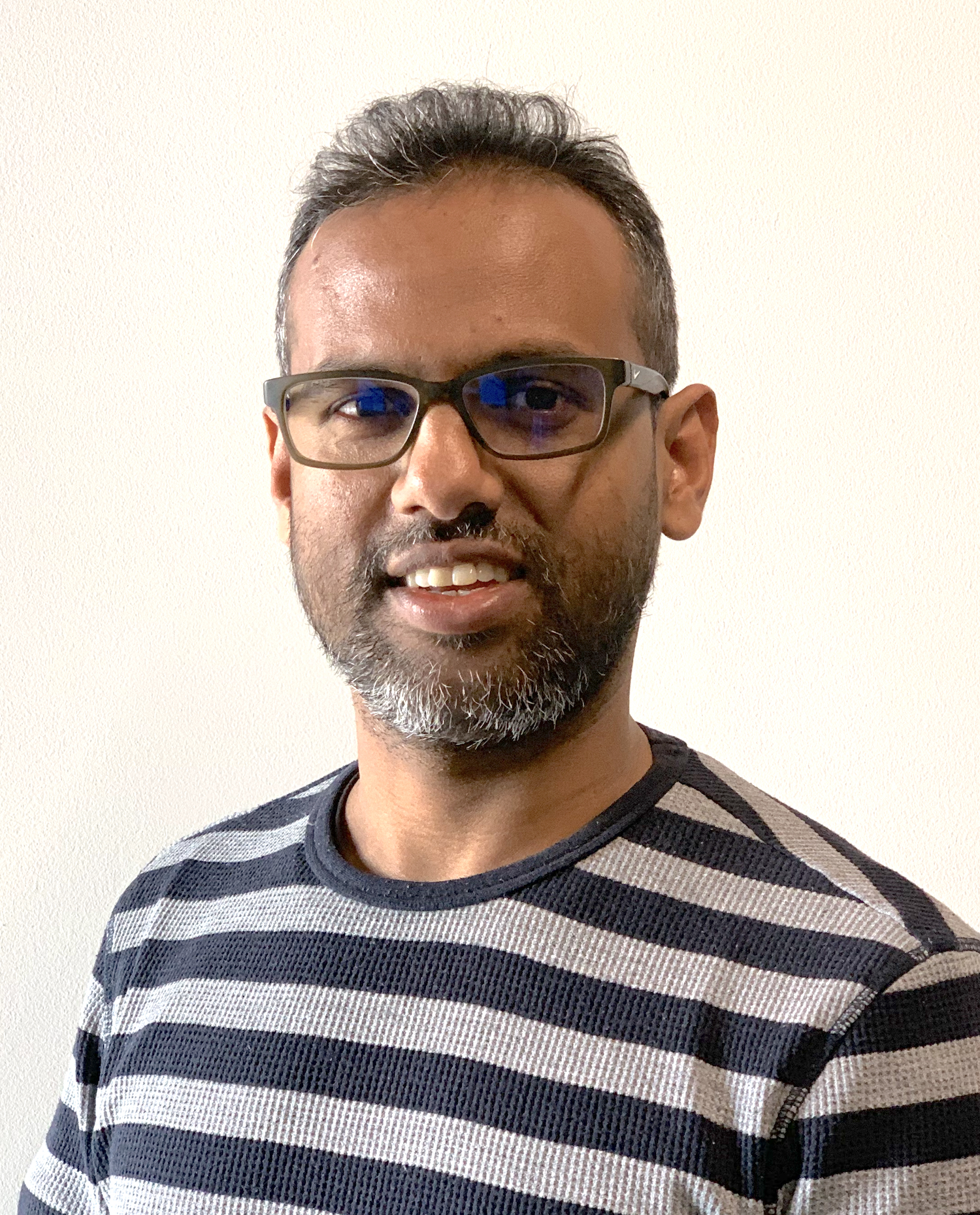}}]{Md Mohaimenuzzaman} is a machine learning software engineer specialising in deep learning in resource limited devices. He obtained a Doctorate in Artificial Intelligence from the Department of Data Science and AI, Faculty of Information Technology at Monash University in Australia. His research centred around deep learning on microcontrollers. Prior to commencing his doctoral studies, he worked as a software engineer for over a decade. In 2007, he earned a bachelor's degree in computer science, and in 2013, he earned a master's degree in the same field.
\end{IEEEbiography}
\vspace{-25pt}
\begin{IEEEbiography}
[{\includegraphics[width=1in,height=1.25in,clip,keepaspectratio]{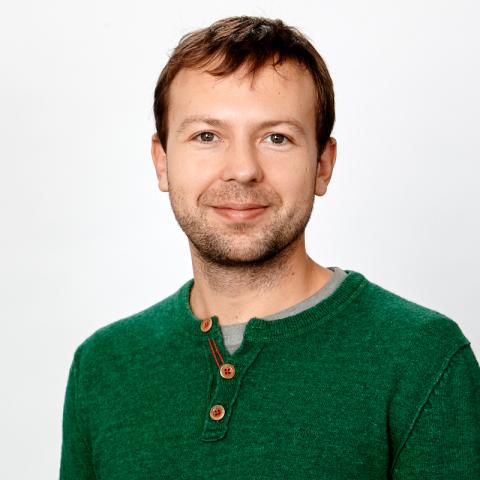}}]{Christoph Bergmeir} is a María Zambrano (Senior) Fellow in the Department of Computer Science and Artificial Intelligence at University of Granada, Spain, and an Adjunct Senior Research Fellow in the Department of Data Science and Artificial Intelligence at Monash University. Before this, he was a Visiting Research Data Scientist at Meta Inc. (formerly Facebook Inc.) in California in the US, and a Senior Lecturer at Monash University. Christoph holds a PhD in Computer Science from the University of Granada, and an M.Sc. degree in Computer Science from the University of Ulm, Germany. 
\end{IEEEbiography}
\vspace{-15pt}
\begin{IEEEbiography}
[{\includegraphics[width=1in,height=1.25in,clip,keepaspectratio]{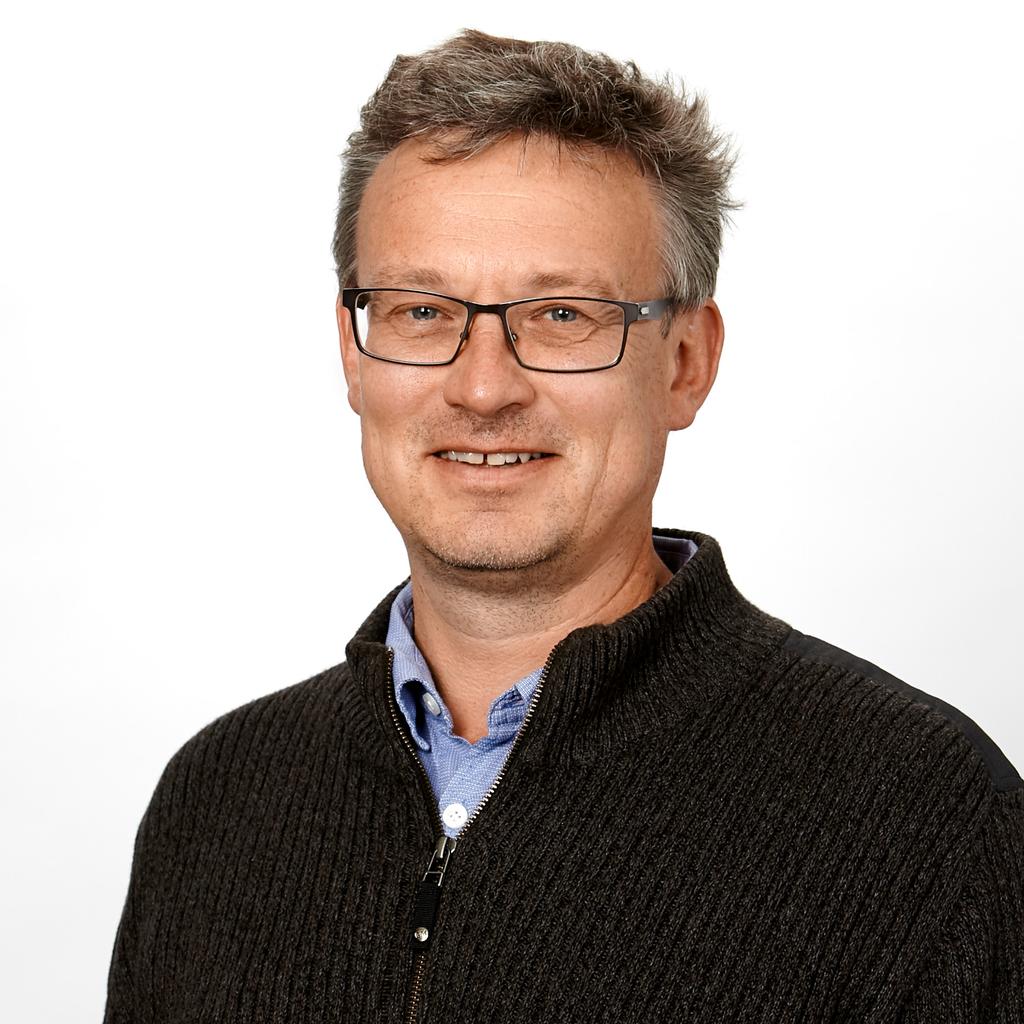}}]{Bernd Meyer} is a Professor in the Department of Data Science and AI, Faculty of Information Technology at Monash University, Australia. He received his PhD in computer science in 1994. Bernd develops mathematical and computational models to explain the collective behavior of social insects, such as bees and ants. He also works on AI-based methods for monitoring animal activity for ecosystem monitoring and automating experiments.
\end{IEEEbiography}

\vfill

\end{document}